\documentclass[twocolumn,times]{aastex63}

\usepackage[normalem]{ulem}         % for \sout
\usepackage{verbatim}
\usepackage{grffile}
\usepackage{jmc_defs190520}
\usepackage{xcolor}

\newcommand{\be}{\begin{eqnarray}}
\newcommand{\ee}{\end{eqnarray}}

\newcommand{\Ftilde}{\widetilde{F}}
\newcommand{\DM}{{\rm DM}}

\newcommand{\EMs}{{\rm EM_s}}

\newcommand{\Gscatt}{G_{\rm scatt}}

\usepackage[utf8]{inputenc}
\usepackage{amsmath}

\newcommand{\DMunit}{pc cm$^{-3}$}

\newcommand{\cnsq}{{\rm C_n^2}}
\newcommand{\linner}{l_{\rm i}}
\newcommand{\louter}{l_{\rm o}}

\providecommand{\DMMW}{\DM_{\rm MW}}

\newcommand{\DMMWh}{\DM_{\rm MW,h}}
\newcommand{\DMMWd}{\DM_{\rm MW,d}}

\providecommand{\DMh}{\DM_{\rm h}}

\newcommand{\Atau}{A_\tau}

\renewcommand{\Gscatt}{G}
\newcommand{\dsl}{d_{\rm sl}}
\newcommand{\dlo}{d_{\rm lo}}
\newcommand{\dso}{d_{\rm so}}

\newcommand{\nud}{\Delta \nu_{\rm d}}

\newcommand{\DMhhat}{\widehat{\DM}_{\rm h}}

\providecommand{\DMigm}{\DM_{\rm igm}}
\providecommand{\DMIGM}{\DM_{\rm igm}}
\providecommand{\DMigmbar}{\overline{\DM}_{\rm igm}}

\providecommand{\figm}{f_{\rm igm}}			% fraction of baryons in IGM

\newcommand{\ACFI}{R_I}
\newcommand{\Dnu}{\Delta\nu}

\newcommand{\FtGunits}{{\rm (pc^2\ km)^{-1/3}}}
\newcommand{\FtG}{\Ftilde G}
\newcommand{\zhat}{{\widehat{z}}}

\begin{document}

\title{The Large Dispersion and Scattering of FRB 20190520B are Dominated by the Host Galaxy}

\author[0000-0002-4941-5333]{Stella Koch Ocker}
\author[0000-0002-4049-1882]{James M. Cordes}
\author[0000-0002-2878-1502]{Shami Chatterjee}
\affiliation{Department of Astronomy and Cornell Center for Astrophysics and Planetary Science, Cornell University, Ithaca, NY 14853, USA}
\author[0000-0001-6651-7799]{Chen-Hui Niu}
\affiliation{National Astronomical Observatories, Chinese Academy of Sciences, Beijing 100101, China}
\author[0000-0003-3010-7661]{Di Li}
\affiliation{National Astronomical Observatories, Chinese Academy of Sciences, Beijing 100101, China}
\affiliation{NAOC-UKZN Computational Astrophysics Centre, University of KwaZulu-Natal, Durban 4000, South Africa}
\author[0000-0002-2885-8485]{James W. McKee}
\affiliation{Canadian Institute for Theoretical Astrophysics, University of Toronto, 60 Saint George Street, Toronto, ON M5S 3H8, Canada}
\author[0000-0002-4119-9963]{Casey J. Law}
\affiliation{Cahill Center for Astronomy and Astrophysics, MC 249-17 California Institute of Technology, Pasadena, CA 91125, USA}
\affiliation{Owens Valley Radio Observatory, California Institute of Technology, 100 Leighton Lane, Big Pine, CA, 93513, USA}
\author[0000-0002-9390-9672]{Chao-Wei Tsai}
\affiliation{National Astronomical Observatories, Chinese Academy of Sciences, Beijing 100101, China}
\author[0000-0001-8057-0633]{Reshma Anna-Thomas}
\affiliation{Department of Physics and Astronomy and the Center for Gravitational Waves and Cosmology, West Virginia University, Morgantown, WV 26506, USA}
\author[0000-0002-4997-045X]{Ju-Mei Yao}
\affiliation{National Astronomical Observatories, Chinese Academy of Sciences, Beijing 100101, China}
\affiliation{Xinjiang Astronomical Observatory, Chinese Academy of Sciences, 150 Science 1-Street, Urumqi, Xinjiang 830011, China}
\author[0000-0001-6804-6513]{Marilyn Cruces}
\affiliation{Max-Planck-Institut für Radioastronomie, Auf dem Hügel 69, D-53121 Bonn, Germany}
\correspondingauthor{Stella Koch Ocker}
\email{sko36@cornell.edu}

% new DMh in obs frame: $903^{+72}_{-111}$
% equivalent in gal frame: $1121^{+89}_{-138}$

\begin{abstract}
The repeating fast radio burst FRB~20190520B is localized to a galaxy at $z=0.241$, much closer than expected given its dispersion measure $\DM=1205\pm4 \ \DMunits$. Here we assess implications of the large DM and scattering observed from FRB~20190520B for the host galaxy's plasma properties. \replaced{Using a sample of 75 bursts detected with the Five-hundred-meter Aperture Spherical radio Telescope, we obtained a mean scattering time $\tau=10.9\pm1.5$ ms at 1.41 GHz, which can be attributed to the host galaxy. The mean scintillation bandwidth of $\nud=0.21\pm0.01$ MHz at 1.41 GHz is consistent with Galactic diffractive interstellar scintillation.}{A sample of 75 bursts detected with the Five-hundred-meter Aperture Spherical radio Telescope shows scattering on two scales: a mean temporal delay $\tau(1.41\ {\rm GHz})=10.9\pm1.5$ ms, which is attributed to the host galaxy, and a mean scintillation bandwidth  $\nud(1.41\ {\rm GHz})=0.21\pm0.01$ MHz, which is attributed to the Milky Way.} 
Balmer line measurements for the host imply an H$\alpha$ emission measure (galaxy frame) $\rm EM_s=620$ pc cm$^{-6}\times(T/10^4 {\rm K})^{0.9}$, implying $\DM_{\rm H\alpha}$ of order the value inferred from the FRB DM budget, $\DMh=1121^{+89}_{-138}$ \DMunit\ for \added{plasma temperatures greater than the typical value $10^4$ K}. Combining $\tau$ and $\DMh$ yields a nominal constraint on the scattering amplification from the host galaxy $\Ftilde G=1.5^{+0.8}_{-0.3} \FtGunits$,  where $\Ftilde$ describes turbulent density fluctuations and $G$ represents the geometric leverage to scattering that depends on the location of the scattering material. For a two-screen scattering geometry where $\tau$ arises from the host galaxy and $\nud$ from the Milky Way, the implied distance between the FRB source and dominant scattering material is $\lesssim100$ pc. The host galaxy scattering and DM contributions support a novel technique for estimating FRB redshifts using the $\tau-\DM$ relation, and are consistent with previous findings that scattering of localized FRBs is largely dominated by plasma within host galaxies and the Milky Way.
\end{abstract}

\section{Introduction}

A substantial fraction of cosmic baryons is in the intergalactic medium (IGM), but they are notoriously difficult to measure with most cosmological observations. Fast radio bursts (FRBs) offer a promising new probe of this baryon content by measuring dispersive propagation delays  caused by ionized media along the line of sight (LOS), including the IGM \citep{2020Natur.581..391M}.  However the contribution to the dispersion measure (DM) from the host galaxies of FRB sources remains one of the largest sources of uncertainty in determining the cosmic baryon fraction of the IGM from FRBs. Recently, the repeating FRB 20190520B (hereafter FRB 190520) was discovered by \citet[][]{2021arXiv211007418N} with the Five-hundred-meter Aperture Spherical radio Telescope (FAST, \citealt{nan11,li19}) and found to have a large $\DM = 1205\pm4$ pc cm$^{-3}$ but a relatively small redshift $z = 0.241$ for its dwarf host galaxy,  J160204.31$-$111718.5 (hereafter referred to as HG190520).  The large implied DM contributed by the host galaxy is an example where the uncertainty in estimating the DM contribution from the IGM is much larger than often assumed. 

The vast majority of known FRBs do not have redshift measurements, and require an inventory of DM contributions from the Milky Way, intervening galaxies, and host galaxies in order to disentangle the DM contribution of the IGM and obtain redshift estimates. For FRB 190520, naive estimates of the DM budget without the host localization would place the source at a redshift $z>1$, demonstrating the significant impact that underestimated host DMs can have on DM-derived distances. These results raise the question of whether the interstellar medium (ISM) of HG190520 contains an anomalously large electron density content, or whether other FRB host galaxy DMs are being systematically underestimated.

\indent It has recently been demonstrated that a combined analysis of FRB scatter broadening along with DM can significantly improve redshift estimates \citep{2021arXiv210801172C}.  %\textcolor{red}{the orginal texts have 'based solely on propagation effects'. But isn't scattering also propagation effect?}
Like dispersion, scattering can occur anywhere along the LOS, but it appears to be dominated by host galaxies, from which it is manifested as temporal broadening  of bursts by a time $\tau$,  and more weakly by the disk of the Milky Way,  as intensity variations with a characteristic frequency scale $\nud$, often called the scintillation bandwidth.
 
Scattering from the ISM of the Milky Way is characterized using observations of the pulsar population, leading to the NE2001 Galactic electron density model \citep{2002astro.ph..7156C}
and YMW16 \citep{2017ApJ...835...29Y}, and in this paper we use NE2001 when necessary. The measured burst scattering time $\tau$ from the host galaxy is used in 
tandem with the $\tau-\DM$ relation to improve the estimated DM contribution of the host galaxy and consequently a propagation-based redshift, $\zhat(\DM, \tau)$. Moreover, combining DM and scattering constraints on the host galaxy ISM (or any other medium along the LOS, such as a galaxy halo) yields information about turbulence-driven density fluctuations in the ionized gas. The statistical properties of density fluctuations can be quantified using the fluctuation parameter $\Ftilde$, which is proportional to $\tau/\DM^2$ \citep[][and references therein]{2021ApJ...911..102O}. It is parameterized as $\Ftilde = \zeta \varepsilon^2 / f (\linner^2 \louter)^{1/3}$  in the context of a medium comprising cloudlets with filling factor $f$, internal and cloud-to-cloud variance quantified by $\zeta$ and $\varepsilon$, and inner and outer scales $\linner, \louter$. 
  
Radio diagnostics are combined with optical imaging and Balmer line spectroscopy to determine or constrain the temperature of ionized gas, which can help discriminate between Milky Way type interstellar media  from gas in extreme conditions in the local environment of FRB sources. 
As such, combining the DM and scattering budgets for a given FRB can not only bolster a redshift estimate for the host galaxy, but can also help characterize the host galaxy ISM, particularly when a host galaxy association yields complementary observations.

\indent In this paper, we seek to distinguish large-scale properties of ionized gas in FRB 190520's host galaxy using the average DM and scattering characteristics of 75 bursts detected by FAST between April and September 2020. Section~\ref{sec:key_obs} summarizes key observations of the FRB and provides measurements of the mean scattering time and scintillation bandwidth. \added{While there is evidence that the scattering time varies between bursts (see Section~\ref{sec:pbf_analysis}), these variations do not appear to follow a systematic trend over time for the burst sample considered here, and instead appear to be stochastic (S.K. Ocker et al., in preparation).} In Section~\ref{sec:host_properties} the DM contribution of the host galaxy is interpreted in tandem with Balmer line measurements and the observed scattering, in order to infer properties of the host galaxy ISM. Section~\ref{sec:taudmz} demonstrates how the construction of a joint scattering-DM budget for this FRB improves its redshift estimation in the absence of a localization. Implications of these results for the plasma properties of HG190520 are discussed further in Section~\ref{sec:discussion}.

We use the following notation to refer to DM contributions from various LOS components: $\DMh$ is the host galaxy DM contribution in the source frame; 
$\DM_{\rm MW}$ is the Milky Way contribution (which we occasionally separate into  halo and disk components); and $\DMIGM$ is the IGM contribution in the observer frame. Similar notation may be used for different LOS contributions to scattering. $\DM_{\rm H\alpha}$ refers to the DM contribution of the host galaxy inferred from H$\alpha$ emission.
 
%\textcolor{red}{Define $\DM_{\rm H\alpha}$ here? It was mentioned in the abstract.}

\section{Key Observations of FRB 190520}\label{sec:key_obs}

\indent FRB 190520 was discovered in the Commensal Radio Astronomy FAST Survey (CRAFTS, \citealt{li18}) with the FAST telescope in drift-scan mode \citep{2020ApJ...895L...6Z}. The four bursts initially discovered in 2019, along with 75 other bursts detected through follow-up tracking observations in 2020, are discussed by \cite{2021arXiv211007418N}, who report a mean $\DM = 1205\pm4$ pc cm$^{-3}$.
Radio imaging with the Karl G. Jansky Very Large Array (VLA) showed the presence of a compact, persistent  radio source (PRS) spatially coincident with the burst source, with a flux density of $202\pm8$ $\mu$Jy at 3 GHz. Follow-up observations at Green Bank Observatory have also revealed extreme rotation measure (RM) variations $\sim 300~\rm rad~m^2~day^{-1}$ over a week-long timespan, indicative of a highly dynamic source environment \citep{2022arXiv220211112A}.

\subsection{Optical Imaging and Spectroscopy}
\label{sec:OIS}

\indent Localization of the FRB enabled  optical and infrared observations  with CFHT/MegaCam, Subaru/MOIRCS, Palomar/DBSP, and Keck/LRIS that  revealed the FRB host to be a dwarf galaxy at a redshift $z = 0.241$. 
The optical spectrum obtained with Keck/LRIS shows H$\alpha$, H$\beta$, $\rm [OIII]4859$\AA, and $\rm [OIII]5007$\AA\ \citep{2021arXiv211007418N} and will be discussed further   in a separate paper (C.W. Tsai et al., in preparation).
 
The Milky Way extinction is estimated to be $E_{B-V} = 0.25$ from the \cite{2011ApJ...737..103S} Galactic dust extinction map, which yields $A_V = 0.76$ assuming $A_V/E_{B-V} = 3.1$. After correction for Milky Way extinction,  the H$\alpha$ and H$\beta$ fluxes are $F_{\rm H\alpha} = 23.9\pm0.3 \times 10^{-17}$ erg cm$^{-2}$ s$^{-1}$ and $F_{\rm H\beta} = 6.2\pm0.3\times10^{-17}$ erg cm$^{-2}$ s$^{-1}$. The H$\alpha$/H$\beta$ line ratio yields an estimate of the intrinsic extinction within the host galaxy equal to 0.8, implying an intrinsic H$\alpha$ flux $F_{\rm H\alpha} = 42.0\pm0.5 \times 10^{-17}$ erg cm$^{-2}$ s$^{-1}$. 

While the host galaxy was unresolved due to atmospheric seeing, we estimate the approximate host galaxy dimensions from the CFHT/MegaCam images to be about 0.5 by 0.5 arcseconds, yielding an H$\alpha$ surface density $S_{\rm H\alpha} \approx 224 \pm 3$ Rayleighs (R) in the source frame, using $z = 0.241$. The error quoted on $S_{\rm H\alpha}$ only accounts for the measurement error of $F_{\rm H\alpha}$, and does not include uncertainty in the size of the host galaxy. The source frame surface density implies an  emission measure in the source frame ($\EMs$),
\begin{equation}\label{eq:EM_vs_T}
\begin{split}
\EMs  &= 
2.75 \ {\rm pc~cm^{-6}} 
T_4^{0.9} S({\rm H\alpha}) \\
&\approx 
616 \pm 7 \ {\rm pc~cm^{-6}}
\times
T_4^{0.9} 
 \left[\frac{S({\rm H}\alpha)}{224 \pm 3 \ {\rm R}}\right] ,
 \end{split}
\end{equation}
where $T_4$ is the temperature in units of $10^4$ K. 

\begin{figure*}[t]
    \centering
    \includegraphics[width=\textwidth]{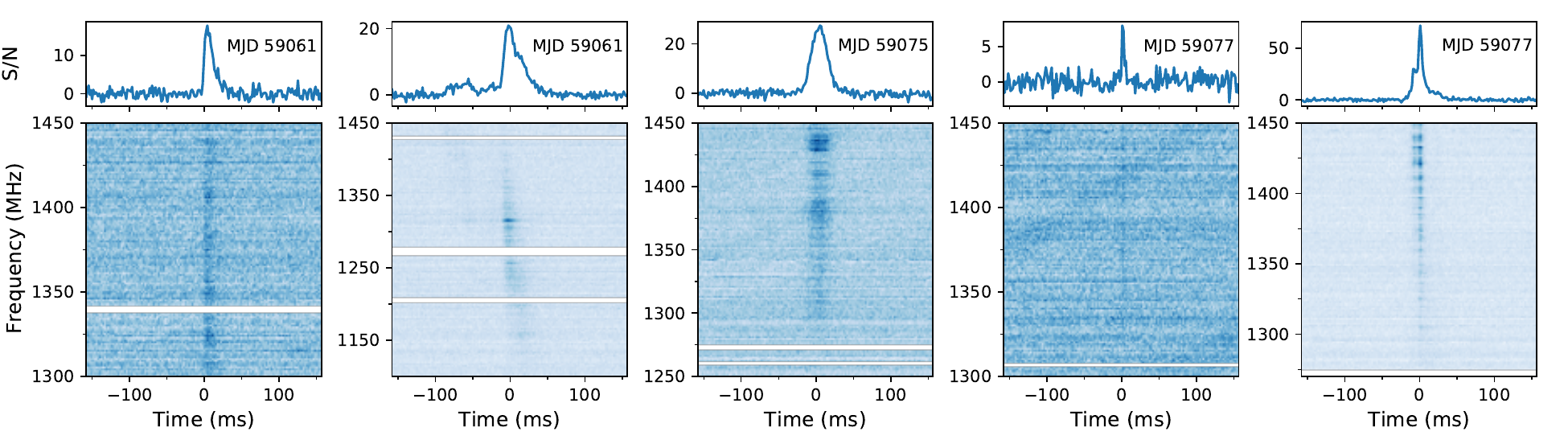}
    \caption{Dynamic spectra of five bursts from FRB 190520 detected by FAST, demonstrating the range of burst properties in frequency and time. The top panels show the \added{frequency-averaged} burst profiles in units of the signal-to-noise ratio (S/N). Masked frequency channels contaminated by RFI are shown in white in the bottom panel.}
    \label{fig:bursts}
\end{figure*}

\subsection{Time-frequency Structure of Radio Bursts}
\label{sec:tfstructure}

We present a detailed analysis of the 75 bursts detected with FAST in 2020 \citep{2021arXiv211007418N} that identifies  temporal broadening of bursts from multipath propagation as well as frequency structure  imposed on burst spectra. We attribute these to scattering in the host galaxy HG190520 and in the Milky Way, respectively, because the pulse broadening is too large to be caused by the IGM or by the halos of either galaxy \added{(both of which are $\lesssim 10\ \mu$s or smaller; \citealt{2013ApJ...776..125M, 2021ApJ...911..102O})} and the observed frequency structure is consistent with expectations based on scintillation measurements of Galactic pulsars. Later we interpret pulse broadening and scintillation together to place an upper bound on the distance of the host scattering region from the FRB source and also to characterize the ISM of the host galaxy. 

The dynamic spectra of five bursts from FRB 190520 are shown in Figure~\ref{fig:bursts} for a 300~ms window centered on each burst and covering \deleted{a selected frequency range} \added{the frequency band in which burst emission is observed,} which is slightly different for each burst. 
Above each dynamic spectrum is the burst profile obtained by averaging over frequency. The bursts have been dedispersed with values of DM that maximize the temporal  structure (as opposed to maximizing signal-to-noise ratio (S/N); \citealt{2019ApJ...876L..23H}).  There are nominal DM variations $\sim 10$ pc cm$^{-3}$ between observing epochs that are highly influenced by burst structure, and it is unclear whether these variations are related to density variations along the LOS or some other process. The average burst DM is $\langle \DM\rangle = 1205\pm 4~\DMunits$ \citep{2021arXiv211007418N}. The FAST digital backend employed a polyphase filterbank that applied a Hamming window to the original voltage data, which were then downsampled to provide the frequency and time resolutions of 0.122~MHz and 96~$\mu s$, respectively. The combined effects of the Hamming window and downsampling yield a resolution function that we estimate to have a width approximately equal to $\Delta \nu_{\rm w} = 0.04$ MHz, \replaced{which we include in our analysis}{which is added in quadrature to the downsampled resolution of 0.122 MHz to approximate the actual frequency resolution of the data for the scintillation bandwidth analysis}. 

The majority of the bursts from FRB 190520 discussed here were detected in the upper part of the FAST observing band, $\sim 1.25$ to $1.45$ GHz, in part because of their emitted spectral shapes but also because substantial radio frequency interference (RFI)  between 1.1 and 1.3 GHz contaminated many of the observations. Three bursts had RFI covering more than $50\%$ of their visible bandwidth and were excluded from the sample, leaving 72 bursts in the following analysis. 

\indent The bursts in Figure~\ref{fig:bursts} display intensity structure in time and frequency that varies substantially from burst to burst. Structure that is intrinsic to the emission process, which may include the drift of intensity islands to later times at lower frequencies (the so-called `sad trombone' effect; e.g. \citealt[][]{2019ApJ...876L..23H}), is modified by pulse broadening that varies strongly with frequency ($\tau \propto \nu^{-x}$) and scintillation intensity frequency structure characterized by the scintillation bandwidth ($\nud \propto \nu^{+x}$). The spectral index is $x = 4.4$ for a Kolmogorov electron density spectrum and for scales in the inertial range between the inner and outer scales. Very strong scattering causes scales smaller than the inner scale to dominate, giving $x \sim 4$, as will other forms of the density spectrum that are steeper than the Kolmogorov form.
%The scintillation bandwidth varies strongly with frequency, $\nud \propto \nu^{+x}$ with $x = 4.4$ for a Kolmogorov electron density spectrum and for scales in the inertial range between the inner and outer scales.  Very strong scattering causes scales smaller than the inner scale to dominate, giving $x \sim 4$, as will other forms of the density spectrum that are steeper than the Kolmogorov form.

\subsubsection{Pulse Broadening Analysis}\label{sec:pbf_analysis}

Burst profiles are the convolution of the emitted burst shape with  a pulse broadening function (PBF), often taken to be a one-sided exponential $\propto \exp(-t/\tau)\Theta(t)$, where $\Theta(t)$ is the Heaviside function (or unit step function).  While pulsar observations indicate the relevance of non-exponential PBFs (e.g. from scattering by Kolmogorov-like density fluctuations  distributed  along the LOS),  the exponential form suffices for our goals here, which require only a characteristic scattering time, $\tau$.

Bursts in Figure~\ref{fig:bursts} show varying degrees of asymmetry that ordinarily would be interpreted as scatter broadening if the broadening time follows the $\tau\propto\nu^{-4}$ scaling. Burst (a) shows plausible scatter broadening whereas bursts (c), (d), and (e) do not.
Burst (b) shows asymmetry but this may be due to drifting (sad-trombone) substructure, as the time-frequency drift of this burst is inconsistent with the $\nu^{-2}$ scaling that would be expected from assuming an incorrect DM. Some of these differences may result from variations in the bursts' spectra, with those more concentrated at higher frequencies expected to show smaller scattering times and {\it vice versa}.  However, not all of the variations in asymmetry seen among the burst sample can be explained this way. 
Measurements of Galactic pulsars show much more consistency in asymmetries from scattering, implying either that the asymmetries evident in Figure~\ref{fig:bursts} are not due to scattering or that the scattering varies substantially between bursts.

In the following we first demonstrate that the timescale $\tau$ for  asymmetries visible in some of the bursts do in fact scale with frequency as expected from scattering (on average).  We then characterize the apparent range of variability of $\tau$ between bursts. 

To assess the presence of scattering, we analyze the shapes of bursts in the Fourier domain by calculating their power spectra \added{in three radio frequency subbands}. The advantage of this approach is that the spectrum is  the product of the  emitted burst spectrum and the PBF spectrum, so the shapes of these factors do not depend on the mean arrival time of the burst (owing to the shift theorem for Fourier analysis). Averaging  bursts in the time domain, by contrast, would be strongly affected by frequency-time drifts that differ between bursts or burst components, \added{particularly if bursts are averaged over the entire frequency band.} 

We compute 1D profiles for the sample of 72 bursts in three radio frequency subbands, $1.05-1.25$ GHz, $1.29-1.37$ GHz, and $1.37-1.45$ GHz, by averaging over frequency and calculating power spectra as the squared magnitude of the \added{fast Fourier transform} (FFT) of each profile. We use a larger bandwidth in the lowest radio frequency subband due to the low number of bursts with emission in this subband, and we omit the $1.25-1.29$ GHz band from all bursts due to strong RFI. The resulting power spectra are averaged over all bursts falling within a given subband: about 15 burst power spectra are averaged in the $1.05-1.25$ GHz subband and about 70 spectra in the upper two subbands. \added{These subbands were chosen based largely on S/N constraints. Future data sets showing large fractions of bursts with frequency-time drift should instead define subbands based on the frequency intervals over which the drift rate is minimized or effectively constant.}

\begin{figure}
    \centering
    \includegraphics[width=0.45\textwidth]{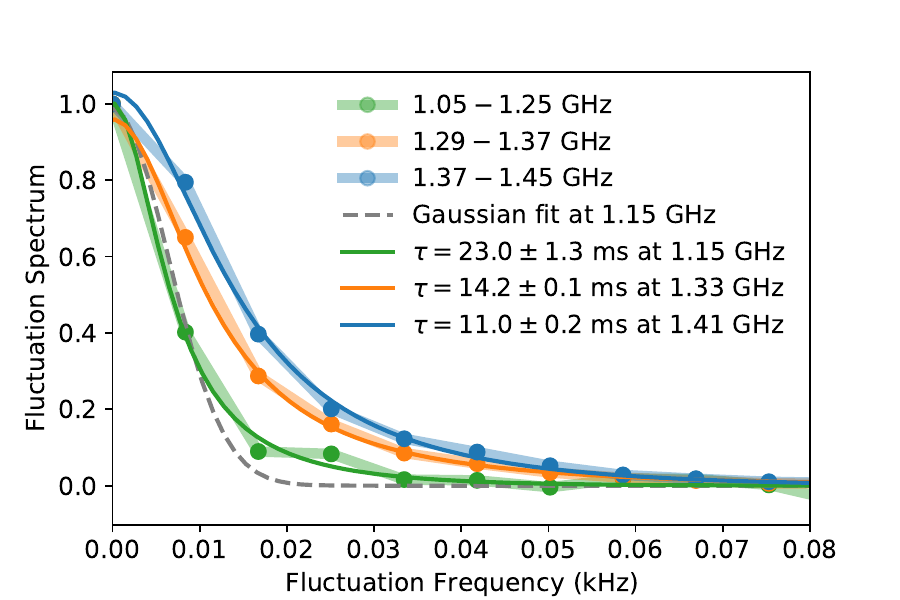}
    \caption{The mean fluctuation spectrum of burst profiles as a function of fluctuation frequency in kHz for three frequency subbands, 1.05-1.25 GHz (light green dotted curve), 1.29-1.37 GHz (light orange dotted curve), and 1.37-1.45 GHz (light blue dotted curve). The fractional error on the mean fluctuation spectrum is of order $1/\sqrt{n}\sim12\%$ for the two upper subbands and $\sim25\%$ for the lowest subband. The dark green, orange, and blue curves show nonlinear least squares fits for a Gaussian pulse convolved with a one-sided exponential PBF in each frequency subband, yielding the scattering times indicated in the legend. The grey dashed line shows a Gaussian pulse fit to the Fourier spectrum at 1.15 GHz.}
    \label{fig:FFTfit}
\end{figure}

Figure~\ref{fig:FFTfit} shows the mean spectra for the three radio-frequency subbands plotted against fluctuation frequency ($f_t$). We interpret their shapes using the product of the  spectrum  for an exponential PBF, 
$S_{\rm PBF}(f_t) \propto \tau/[1 + (2\pi \tau f_t)^2]$ with  a Gaussian-shaped power spectrum,  $S_{\rm G} \propto \exp(-(2\pi\sigma f_t)^2$ for an assumed Gaussian burst shape for emitted profiles using $\sigma$ as the half-width at $e^{-1/2}$.

Figure~\ref{fig:FFTfit} also shows the fit of the product $S_{\rm PBF}(f_t) S_G(f_t)$ to the mean burst power spectrum centered on 1.41 GHz, which yields $\tau = 11.0\pm0.2$~ms and a full-width-at-half-maximum of the Gaussian component $\rm FWHM = 5.7 \pm 0.5$~ms. An independent fit at 1.33~GHz yields $\tau = 14.2 \pm 0.1$~ms and $\rm FWHM = 4.2 \pm 0.2$~ms, and a fit at 1.15~GHz yields $\tau = 23.0 \pm 1.3$~ms and $\rm FWHM = 8.7 \pm 3.3$~ms. These three scattering times are consistent with a $\nu^{-4}$ frequency scaling to within 1 standard deviation. 
The difference between the Gaussian FWHM fit within each subband is largely due to different bursts occupying different parts of the radio frequency band. A Gaussian pulse spectrum fit at 1.15 GHz is also shown in Figure~\ref{fig:FFTfit} for comparison. A simultaneous fit to all three frequency subbands assuming $\tau \propto \nu^{-4}$ yields $\tau = 10.9 \pm 1.5$ ms at 1.41 GHz with a reduced chi-square $\bar{\chi}^2 = 0.64$. Assuming $\tau \propto \nu^{-4.4}$ yields similar results, $\tau = 10.4 \pm 1.3$ ms at 1.41 GHz with $\bar{\chi}^2 = 0.7$. Due to the slightly smaller $\bar{\chi}^2$ we adopt the former value, $\tau = 10.9 \pm 1.5$ ms at 1.41 GHz, as the mean scattering time for the remainder of our analysis.
%\textcolor{red}{explain why -4.4 here}

Our scattering time estimates are about 80\% larger than those in
\cite{2021arXiv211007418N}, who report  $\tau = 9.8 \pm 2$ ms at 1.25 GHz, equivalent to $\tau\approx 6\pm1$ ms at 1.41 GHz ($\approx 2\sigma$ discrepancy). This scattering time was also based on a burst shape model comprising an exponential PBF convolved with a Gaussian function, but fitting was done in the time domain, rather than the frequency domain, on a smaller number of bursts that showed apparent scatter broadening.   Another difference is that the fits were done on burst shapes obtained by integrating over the entire frequency range rather than in subbands.   The observed tendency for this set of bursts to be stronger at higher frequencies in the FAST band, combined with the $\tau\propto \nu^{-4}$ scaling, is likely to have resulted in smaller scattering times estimated from this procedure.  

While the aggregate sample of bursts has a mean scattering time $\tau = 10.9 \pm 1.5$ ms at 1.41 GHz, there is preliminary evidence that $\tau$ may fluctuate from burst to burst. While some bursts with $\rm S/N>10$ show frequency-dependent temporal asymmetries consistent with scattering timescales $\sim 10$ ms at 1.4 GHz, two bursts in the sample are symmetric across the entire radio frequency band and have FWHM $\leq 7$ ms, suggesting that $\tau$ may vary by at least $\sim 3$ ms. However, many of the bursts have too low $\rm S/N$ to distinguish between intrinsic spectral variations and scattering on an individual basis. A more detailed analysis and interpretation of these apparent scattering variations will be discussed in a separate paper.

\subsubsection{Frequency Structure in Burst Spectra}

Observed burst spectra consist of emitted spectral shapes modified by multipath propagation. Emitted bursts are consistent with amplitude modulated shot noise, where shots of $\sim$ns duration determine the overall spectral shape while also, in concert with the modulations, inducing frequency structure with characteristic frequency scales equal to the reciprocals of characteristic burst widths  \citep[e.g.][and references therein]{2022NatAs...6..393N}.  Bursts with $\sim$ms widths produce kHz structure, but $\sim$MHz frequency scales can be produced if there is substructure on microsecond scales. We refer to this frequency structure as `self-noise'.   
 
The spectral modulation from self-noise is 100\% (i.e. RMS intensity = mean intensity).   In this picture, drifting spectral islands  are part of the amplitude modulation and modulated shots determine the center frequencies and spectral widths of the islands. An alternative view is that spectral islands are extrinsically produced by plasma lensing but the systematic trend for lower-frequency islands to arrive later is not naturally produced by lensing \citep[e.g.][]{2017ApJ...842...35C}. 

Diffractive interstellar scintillation (DISS) from multipath scattering in the ISM of the Galaxy also  produces 100\% intensity variations {\it vs.} both  time and frequency for a point source  in the strong-scintillation regime \citep[e.g.][]{ric90}.   The short durations of FRBs imply that DISS will only be identifiable in the frequency domain because scintillation timescales are generally much larger than burst durations. 
The characteristic frequency scale of DISS, the scintillation bandwidth $\nud$, is typically estimated as the half-width-at-half-maximum (HWHM)  of the intensity autocorrelation function (ACF). %The scintillation bandwidth varies strongly with frequency, $\nud \propto \nu^{+x}$ with $x = 4.4$ for a Kolmogorov electron density spectrum and for scales in the inertial range between the inner and outer scales.  Very strong scattering causes scales smaller than the inner scale to dominate, giving $x \sim 4$, as will other forms of the density spectrum that are steeper than the Kolmogorov form. 
While self-noise and DISS share similar statistics,  the strong, characteristic  frequency dependence of $\nud$ from DISS resulting from scattering is likely distinct from that for self-noise frequency structure.

\subsubsection{Scintillation Frequency Structure Analysis}\label{sec:nud_obs}

Galactic DISS has been measured in a fairly small sample of FRBs thus far, including FRB 20110523A \citep{2015Natur.528..523M}, FRB 20121102A (hereafter FRB 121102; \citealt{2019ApJ...876L..23H}), FRB 20180916B \citep{2020Natur.577..190M}, FRB 20200120E \citep{2022NatAs...6..393N}, and several FRBs in CHIME/FRB Catalog 1\footnote{https://www.chime-frb.ca/catalog} \citep{2021RNAAS...5..271S}. Successful detection of Galactic DISS depends on two main factors: sufficient frequency resolution to resolve $\nud$ and small-enough extragalactic scattering so that  the wavefronts incident on the Galaxy have sufficient  spatial coherence. 

At 1~GHz $\nud$ is generally predicted to be $\lesssim5$ MHz using the NE2001 model \citep[][Figure 6]{2019ARA&A..57..417C} and for FRB~190520 ($l = -0.33^\circ$, $b = 29.91^\circ$) the prediction is $\nud \sim 0.5$~MHz.  The frequency sampling of the FAST data $\Delta\nu_{\rm s} = 0.122~$MHz formally resolves this nominal DISS bandwidth but, as shown below, the measured $\nud$ is smaller than the predicted value, requiring special attention to the actual frequency resolution, which is slightly larger than $\Delta\nu_{\rm s}$.  This resolution  quenches the $\lesssim$~kHz frequency structure from self-noise but any burst substructure smaller than about $10~\mu$s may contribute to burst spectra. 

The spatial-coherence requirement implies that the angular diameter, and thus the scattering time $\tau$,  produced by extragalactic scattering is sufficiently small to allow fully modulated Galactic DISS to occur.    The angular diameter depends on the distance of the  extragalactic scattering screen from the FRB source, so the occurrence of DISS implies an upper bound on that  distance (Section~\ref{sec:Lx};  \citealt{2019ARA&A..57..417C}).

\begin{figure}
    \centering
    \includegraphics[width=0.48\textwidth]{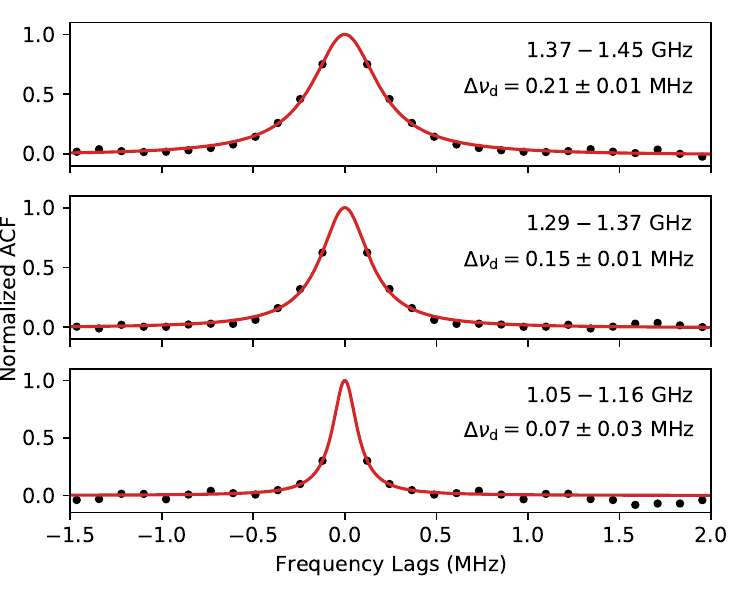}
    \caption{Mean autocorrelation function (ACF) for all bursts within three frequency subbands: $1.37-1.45$ GHz (top), $1.29-1.37$ GHz (middle), and $1.05-1.16$ GHz (bottom). The mean ACF is calculated by dividing the bursts into subbands and then calculating the ACF for each burst, before averaging all of the burst ACFs within each subband. The ACFs shown here have been normalized to a maximum of one after averaging, and are only shown for frequency lags between $-1.5$ and $1.5$ MHz. The black points correspond to the measured ACFs, and the red lines correspond to nonlinear least squares fits for a modified Lorentzian model with a scintillation bandwidth $\nud$ indicated in each panel. The errors on the measured ACF values are smaller than the size of the black points shown.}
    \label{fig:acf_fit}
\end{figure}

Operationally, the scintillation bandwidth is estimated as the HWHM of the intensity ACF  after due allowance for a narrow spike at zero lag from radiometer noise.   For $x=4$ and a thin screen, the theoretical form for the ACF is Lorentzian \citep[e.g.][]{1998ApJ...505..928G}, 
\be
\ACFI(\Dnu) = \frac{(\nud)^2  + (\Dnu_{\rm w})^2}{(\nud)^2  + (\Dnu_{\rm w})^2 + (\Dnu)^2 },
\ee
where we have included an extra term $(\Dnu_{\rm w})^2$ to account approximately for the spectral window function used in the data acquisition, which synthesizes a polyphase filterbank that approximates a Hamming window. 

Our analysis relies on the strong frequency dependence of $\nud$ to identify DISS.  Similar to pulse broadening, most of the burst spectra discussed here lack sufficient S/N to fit for a scintillation bandwidth individually; instead, the burst ACFs are calculated within frequency subbands and then averaged to produce a mean ACF. We adopt slightly different subband divisions to calculate the scintillation bandwidth: $1.05-1.16$ GHz, $1.29-1.37$ GHz, and $1.37-1.45$ GHz; not only are these subbands generally free of RFI, but they also yield mean burst profiles that are contiguously sampled in frequency and subsequently have consistent lag spacing in the ACF, producing  uniform sampling in the average ACFs. 

After each burst in the 72-burst sample is divided into subbands, the on-burst spectrum is averaged in time. A linear fit to the off-burst noise spectrum is subtracted from the on-burst spectrum before calculating the burst ACF. All burst ACFs with power in a given subband are then averaged within that subband to produce a single, mean ACF.
%All of the burst ACFs are then averaged within each subband to produce a single, mean ACF per subband. 
The mean ACFs for each subband are shown in Figure~\ref{fig:acf_fit}, along with nonlinear least squares fits of the modified Lorentzian model. The zero-lag noise spike is excluded, and the amplitude of the Lorentzian is left as a free parameter within each subband. 

We find $\nud = 0.07\pm0.03$ MHz at 1.105 GHz, $\nud = 0.15\pm0.01$ MHz at 1.33 GHz, and $\nud = 0.21\pm0.01$ MHz at 1.41 GHz. Together these imply  $\nud \propto \nu^{4.7\pm0.5}$ from a least-squares fit to the exponent, which is consistent with the frequency scaling expected for inertial-range Kolmogorov turbulence ($\nu^{4.4}$), and re-scaling the best-fit value of $\nud$ at 1.41 GHz yields $\nud\approx0.05$ MHz at 1 GHz. Alternatively, fixing the spectral index to $4.4$ and fitting for $\nud$ across all three subbands simultaneously yields a best-fit value $\nud = 0.052\pm0.007$ at 1 GHz. \replaced{Both of these fitting approaches are  consistent and demonstrate that the estimated decorrelation bandwidths are consistent with DISS.}{These fitting approaches are consistent with each other and with the scintillation bandwidth independently inferred from GBT observations at 5 GHz by \cite{2022arXiv220211112A}, demonstrating that the estimated decorrelation bandwidths are consistent with DISS.} Because the zero-lag noise spike in the ACF is large due to the generally low signal-to-noise ratios of the bursts and residual contributions from self-noise, we are unable to estimate the intensity modulation index reliably.  Nonetheless, burst self-noise is not expected to be frequency dependent in the same way as DISS, so the lack of a constraint on the modulation index does not impact our interpretation of the ACF.

The scintillation bandwidth predicted by NE2001 along the FRB LOS is $\nud \approx 0.5$ MHz at 1 GHz, almost 10 times larger than the measured value. While a number of FRBs with published scintillation bandwidths are broadly consistent with the NE2001 predictions \citep[e.g.][]{2021ApJ...911..102O, 2021RNAAS...5..271S}, significant deviation from the NE2001 prediction has been observed in at least one other case, FRB 20201124A, and may be related to localized density structure that is not incorporated in the model \citep{2022MNRAS.509.3172M}. Although FRB 190520 lies at a high Galactic latitude ($b = 30^\circ$), its longitude of $-0.33^\circ$ suggests that extra ionized gas associated with the Galactic Center could induce more Galactic scattering (and hence a smaller $\nud$) than expected. Future calibration of electron density models may benefit from considering these discrepancies.

Recently, giant pulses (GPs) from pulsars have been proposed as physical analogs for FRBs due to shared characteristics, especially their short durations, spectral luminosities, and complex time-frequency structure \citep{2016MNRAS.457..232C, 2019ApJ...876L..23H}. The association of the Galactic FRB 20200428 with a coincident X-ray burst from the magnetar SGR 1935+2154 \citep{brb+20} as well as the detection of radio-GP-like emission from the magnetar XTE J1810$-$197 \citep{crd+22} intriguingly suggests a link between GPs, magnetar bursts, and FRBs \citep{2022NatAs...6..393N}.
\deleted{CHIME} Observations with high fractional bandwidth \added{at the Algonquin Radio Observatory} of the Crab pulsar revealed a banding effect \added{on frequency scales of 20 to 50 MHz} in the spectra of GPs very similar to that seen in FRBs. The banding is intrinsic to the GP emission and shifts within the length of the scattering tail \citep{2021ApJ...920...38B}, and may be explained by highly relativistic plasma traveling outward from the light cylinder of the pulsar. Although the emission mechanisms of GPs and FRBs are likely not the same, if the FRB emission mechanism arises from highly relativistic plasma (as in some theoretical predictions; \citealp{lyu21}), such structure might be present in FRB spectra. While we do not find strong evidence for similar frequency banding in bursts from FRB 190520 that cannot be explained solely by scintillation, we note that the S/N of most bursts is insufficient to identify such an effect and that the fractional bandwidth of our data set is not large. \added{Future work may be able to resolve this banding effect if intensity modulations are observed with a frequency dependence that does not follow the strong inverse frequency scaling ($\sim \nu^{-4}$) expected for DISS.}

\section{Properties of the Host Galaxy}\label{sec:host_properties}

 Here we combine  constraints on the dispersion measure $\DMh$ of the host galaxy with the estimated emission measure and scattering measurements to deduce properties of the host galaxy ISM. 
 
\subsection{DM Inventory Analysis}\label{sec:DMinventory}

We summarize briefly the analysis reported in \citet[][]{2021arXiv211007418N} that disentangles the  contributions to DM from the MW, the IGM, and the host galaxy,
\be
\DM = \DMMW + \DMIGM(\zh) + \DMh / (1+\zh),
\ee     
where we separately discuss the disk and halo components of the MW,
$\DMMW = \DMMWd + \DMMWh$, while we lump together  all DM components of the host galaxy and define
$\DMh$ to be the dispersion measure in the galaxy frame (rather than observer's frame).    Our analysis follows that in \citet[][]{2021arXiv210801172C}:
\begin{enumerate}
\item The  NE2001 model is used to estimate $\DMMWd = 60~\DMunits$ and a $\pm 20$\% uncertainty is incorporated with a flat distribution;
\item The MW halo contribution is modeled as a flat distribution in the interval $[25, 80]~\DMunits$;
\item The combined trapezoidal disk and halo distribution gives $\DMMW = 113 \pm 17~\DMunits$ where the
uncertainty is simply the RMS DM. 
\item The contribution from the IGM is estimated using a log-normal distribution with mean  %\textcolor{red}{how about give the log-normal function here?}
$\DMigmbar(\zh)$, and RMS  $\DMigmsig(\zh) = (\DMigmbar(\zh) \DM_{\rm c})^{1/2}$ where $\DM_{\rm c} = 50~\DMunits$ and
\begin{multline}
 \DMigmbar(\zh) \approx  978\,\DMunits\,\\ \times \figm \int_0^{\zh}d\zp\, \frac{(1+\zp)}{E(\zp)}.
\label{eq:DMIGMbar}
\end{multline}
Here  $E(z) = [\Omega_{\rm m} (1+z)^3 + 1 - \Omega_{\rm m}]^{1/2}$ for a flat $\Lambda$CDM universe with a matter density $\Omega_{\rm m}$ and $\figm$ is  the fraction of baryons in the ionized IGM, for a constant assumed value of the baryon density $\Omega_{\rm b}$. Constants were evaluated from the Planck 2018 analysis \citep{2020AA...641A...6P} implemented in {\tt Astropy} \citep{2013AA...558A..33A, 2018AJ....156..123A}.  
\item The posterior \added{probability distribution function} (PDF) for $\DMh$ is calculated using a flat prior and by marginalizing 
$\DMh = (1 + \zh) [\DM - \DMMWd - \DMMWh - \DMigm(\zh)]$ over the distributions  just described above.
No uncertainties were included for the measured redshift or DM.   We used a baryonic fraction
$\figm = 0.85$ as the best fit value found in \citet[][]{2021arXiv210801172C}, although we also varied $\figm$ between 0.4 and 1.2 in our analysis, where  the $\figm  = 1.2$ case allows for the possibility that FRBs are found in overdense regions with an effective value of  $\figm$ exceeding the true cosmological average.  
\end{enumerate}
The calculated posterior yields a median and 68\% probable interval, 
$\DMh = 1121^{+89}_{-138}~\DMunits$ (host frame)
or $\DMh^{(\rm obs)} = 903^{+72}_{-111}~\DMunits$ 
in the observer's frame.  The corresponding IGM contribution
is  $\DMigm = 195^{+110}_{-70}~\DMunits$, which also includes uncertainties in the MW contribution along with cosmic variance of the IGM. 

\begin{figure}
    \centering
    \includegraphics[width=\linewidth]{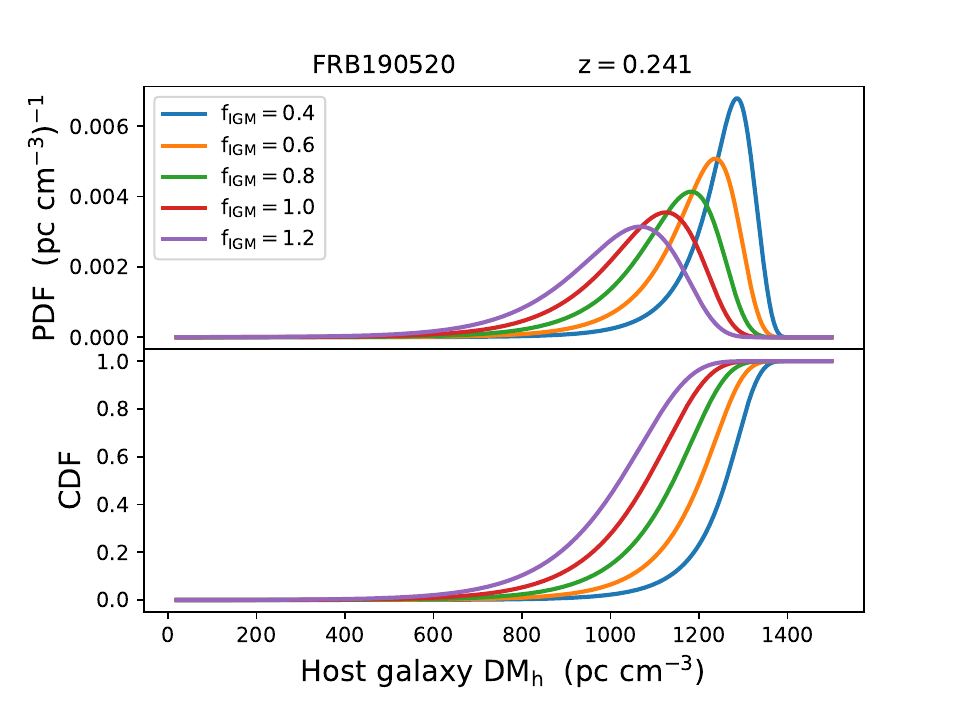}
    \caption{Posterior PDF and CDF for $\DMh$ (in the host galaxy's frame) for five values of the IGM's baryonic fraction $\figm$. PDFs are normalized to unit maximum.}
    \label{fig:DMh_pdf}
\end{figure}

Figure~\ref{fig:DMh_pdf} shows posterior PDFs and \added{cumulative distribution functions} (CDFs) for $\DMh$ for five values of $\figm$ that bracket the nominal value $\figm = 0.85$ used to report our results.   Even for the largest value of $\figm = 1.2$,  the lower bound (84\% probability) exceeds 850~$\DMunits$ (or 682~$\DMunits$ in the observer's frame), far larger than the maximum of the $68\%$ probable range for the IGM contribution (413~$\DMunits$), again for $\figm=1.2$.  

\subsection{Joint Likelihood Analysis of Dispersion Measure and Balmer Lines} \label{sec:jointDMHalpha}

The source-frame EM reported in \S\ref{sec:OIS} can be used to estimate a model-dependent \DM\ from the host galaxy that probes the portion of the LOS along which  H$\alpha$ emitting gas is prevalent. {\em A priori}, the path length probed by the observed H$\alpha$ may be longer than that probed by dispersed bursts if the FRB source is located partway through the H$\alpha$ emitting gas;
however, the H$\alpha$ path length could instead be smaller if there are significant amounts of ionized gas that emit little H$\alpha$. \added{\cite{2021arXiv211007418N} used the observed H$\alpha$ emission to estimate a range of DM values between 230 and 650 pc cm$^{-3}$ (observer frame). Here we expand that analysis by statistically characterizing the H$\alpha$-inferred DM using maximum likelihood estimation.}

The equivalent DM contribution from H$\alpha$ emitting gas is related to EM by 
\begin{equation}\label{eq:EMDM}
    {\rm EM} = \frac{\zeta(1+\epsilon^2)}{f} \frac{{\rm DM}^2}{L}
\end{equation}
\citep[][and references therein]{2017ApJ...834L...7T, 2020ApJ...897..124O}, where $\zeta$, $\epsilon$, and $f$ are parameters in the ionized cloudlet model that characterize cloud-to-cloud  and intracloud density fluctuations and the cloud filling factor, respectively, and $L$ is the path length through the gas. FRB~190520 is offset from the host galaxy center by about $1.3^{\prime\prime}$ \citep{2021arXiv211007418N}, or about 5 kpc, which gives a sense of the scale of path lengths that might reasonably be sampled by the FRB; that being said, the FRB DM only traces a portion of the H$\alpha$ emitting gas. Joint constraints on $\rm EM$ and $\DM$ from the H$\alpha$ emission and the DM inventory can therefore, at least in theory, constrain the gas temperature and $\zeta(1+\epsilon^2)/fL$, but in the absence of independent measurements of $T$ and $\zeta(1+\epsilon^2)/fL$ we must make assumptions about the reasonable ranges of these parameters. 

\indent We first demonstrate the relationship between $\DMh$ and temperature by constructing a likelihood function for $T_4$ using the observed H$\alpha$ surface density and physically motivated priors for $L$, $\zeta(1+\epsilon^2)/f$, and $T_4$ \added{described below}. For $L$, we adopt a log-normal prior with a mean of 2 kpc and standard deviation of 4.5 kpc, which is motivated by the physical scale of the host galaxy \added{\citep{2021arXiv211007418N}}. In the ionized cloudlet model, $\zeta\geq1$, $0\leq \epsilon^2 \leq1$, and $0\leq f \leq 1$ \added{\citep{1991Natur.354..121C, 2002astro.ph..7156C}}. For warm ionized gas ($T_4\sim1$) $f\sim0.1$, whereas for hot ionized gas ($T_4\sim100$) $f$ can be much larger \added{\citep{2011piim.book.....D}}. We therefore adopt a flat prior on the composite parameter $A = \zeta(1+\epsilon^2)/f$ restricted to the range [1, 50] (which encapsulates a range of cases between $f \sim 1$, $\zeta \sim 1$, $\epsilon^2 \ll 1$ and $f \ll 1$, $\zeta \sim \epsilon^2 \sim 1$).  H$\alpha$ emission is typically observed between temperatures of about 5,000 to 16,000 K \citep{2011piim.book.....D}, so we adopt a log-normal prior on $T_4$ with a mean of $1.5$ and a standard deviation of $5$. The H$\alpha$ surface density in the source frame is also given a Gaussian prior with a mean and standard deviation set by the values inferred from the observed H$\alpha$ luminosity, $S_{\rm H\alpha} = 224\pm3$ Rayleighs. For $\DMh$, we initially adopt a flat prior restricted to the range [100, 2000] pc cm$^{-3}$, so that we can explicitly show how $\DMh$ scales with $T_4$ for the observed $S_{\rm H\alpha}$. We then use a numerical grid search to calculate the likelihood function for $T_4$ as $p(T_4|\DMh) \sim \int {\rm d}L {\rm d}A f(L) f(A) \delta(T_4 - g(L,A)$, where $g(L,A)$ is the function relating $T_4$, $S_{\rm H\alpha}$, $\DMh$, and $A$ based on Equations~\ref{eq:EM_vs_T} and~\ref{eq:EMDM}. 

\begin{figure}
    \centering
    \includegraphics[width=\linewidth]{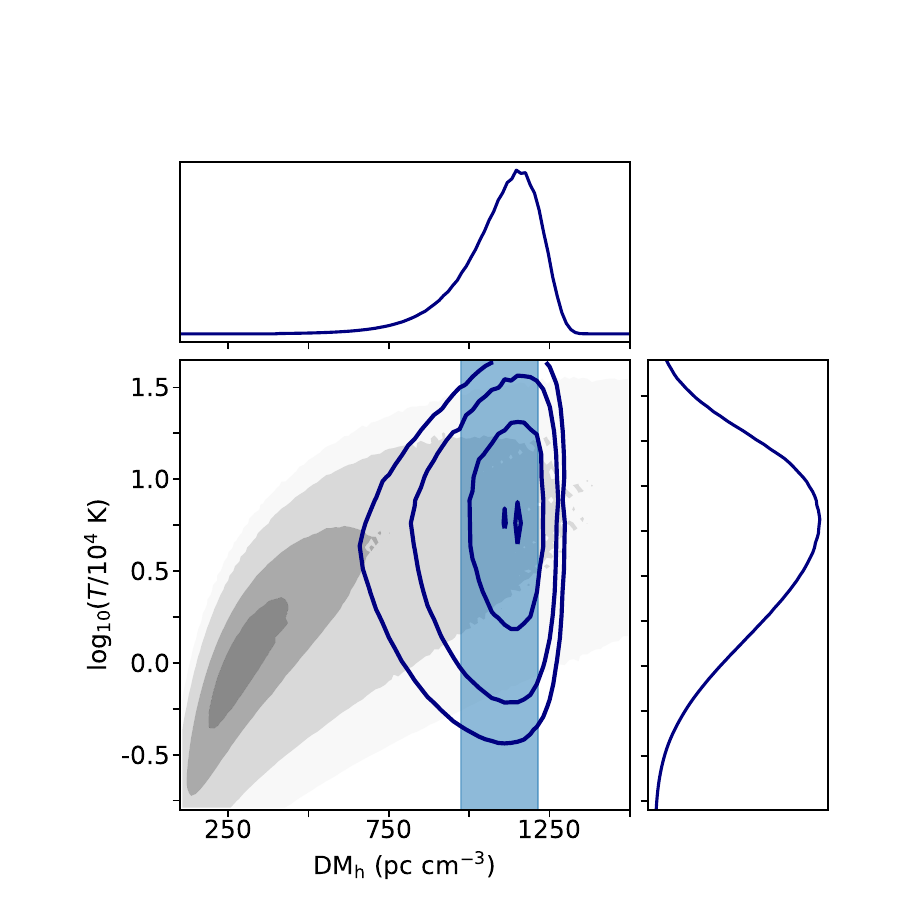}
    \caption{Joint constraints on the temperature and DM contribution of ionized gas in the host galaxy frame of FRB 190520, assuming that the observed H$\alpha$ emission traces the same gas responsible for the host galaxy $\DMh$. The gray shaded contours show the probability density for temperature in units of $10^4$ K vs. $\DMh$, assuming log-normal priors on the path length through the gas and the temperature, flat priors on $\DMh$ and the density fluctuation statistics $\zeta(1+\epsilon^2)/f$, and a Gaussian prior on the H$\alpha$ surface density based on the measured H$\alpha$ luminosity (see Section~\ref{sec:jointDMHalpha}). The blue shaded region shows the $68\%$ probable range for the independently constrained host galaxy $\DMh$ based on the FRB's DM inventory and redshift (see Section~\ref{sec:DMinventory}). The blue contours show the results of a joint likelihood analysis of the gray and blue shaded probability distributions, and the side panels show the marginalized probability distributions for $T$ and $\DMh$ from this joint likelihood analysis. Contour levels correspond to $5\%$, $15\%$, $50\%$, and $85\%$ of the maximum likelihood. }
    \label{fig:T_vs_DM}
\end{figure}

\indent The resulting likelihood function for $T_4$ vs. $\DMh$ is shown in Figure~\ref{fig:T_vs_DM}, along with the range of $\DMh$ that is independently constrained by the DM budget in Section~\ref{sec:DMinventory}. Previous studies \added{\citep[e.g.][]{2011ApJ...736...83H,2017ApJ...834L...7T, 2021ApJ...911..121P}} typically assumed $T_4\approx1$ to convert H$\alpha$ EM to DM, but in this case adopting $T_4\approx1$ yields $\DMh\approx300$ pc cm$^{-3}$, less than half the value inferred from the DM budget, assuming the prior on $A$. In order for the H$\alpha$ emission to explain the FRB's $\DMh$, $T_4$ needs to be almost an order of magnitude larger. A numerical joint likelihood estimate of $p(T_4|\DMh)$ and the PDF for $\DMh$ calculated in Section~\ref{sec:DMinventory} yields $T = 5^{+10}_{-4}\times10^4$ K. This constraint on $T$ assumes that the entire H$\alpha$ EM is attributable to the gas responsible for $\DMh$, but the FRB LOS likely only probes a fraction of the H$\alpha$ emission observed from the entire galaxy. As such, this constraint on $T$ could be regarded as a lower bound, because the temperature would need to be even larger for the FRB $\DMh$ to account for only a fraction of the total H$\alpha$ EM. 

\indent There are a few scenarios that may make the FRB DM budget consistent with the H$\alpha$ emission observed from the host galaxy. One scenario is that the temperature of gas sampled by the FRB is higher than typical for warm, H$\alpha$ emitting gas. Another is that the density fluctuation parameter $A$ is significantly different than our nominal assumptions, which may be the case if turbulence in the host galaxy is significantly different than in the Milky Way. Yet another alternative is that the H$\alpha$ EM conforms to $T_4\sim1$ and $f\sim0.1$, but the total $\DMh$ includes contributions from other (fully) ionized gas in the host galaxy. Any one (or a combination) of these scenarios could explain both the observed H$\alpha$ luminosity and the $\DMh$ implied by the FRB DM budget, but determining which scenario is the most plausible requires additional information about the properties of the host galaxy across multiple phases of its ISM. \added{While an additional galaxy disk intervening the LOS could also make the FRB DM budget consistent with the host H$\alpha$ emission, the scattering in this scenario would be significantly larger than the scattering that is observed, and the optical neighborhood of the host galaxy does not appear to show relevant foreground objects \citep{2021arXiv211007418N}.}

\subsection{Scattering in the Host Galaxy}\label{sec:Lx}

In \S\ref{sec:tfstructure} we established that radio scattering of bursts from FRB~190520  is manifested in two ways: through intensity scintillations from scattering in the MW and pulse broadening from scattering in the host galaxy.    Extragalactic scattering can attenuate MW scintillations by reducing the coherence length of the radio waves incident on the MW, $\lc \simeq \lambda / 2\pi\thetaXo$,  below that needed to produce 100\% intensity modulations, where $\thetaXo$ is the observed angular size of the scattered source.  Here we present a brief summary and defer a more detailed analysis to a paper in preparation.       

The pulse broadening time from an extragalactic screen is $\tauX =  (\thetaXo^2/c) (\dso \dlo / \dsl)$, where $\dso = $~source-observer distance, $\dlo = $~scattering-layer-observer distance, and $\dsl$ is the distance of the scattering layer from the source.    (Often the expression for $\tauX$ would have a factor $1/2c$ rather than $1/c$, but when $\thetaXo$ is taken as the RMS image size of a circular image in one dimension, $\tauX \propto 2\thetaXo^2$, leading to our expression.)
Using $\dlo / \dso \to $1 for a distant FRB source and a scattering screen in the host galaxy at  $\dsl \equiv \LX$  we have $\tauX \sim  (\thetaXo^2/c) (\dso^2/\LX)$.  
The coherence length of the scattered waves is  then  
$\lc \simeq (\lambda \dso) / (2\pi \sqrt{c\tauX\LX})$.     

The required minimum coherence length is the size of the scattering cone $\lcone$ from Galactic scattering projected onto the Galactic scattering screen.  Using analogous definitions,  the observed scattering angle from Galactic scattering is $\thetaGo \simeq  \sqrt{c\tauG/\LG}$ and 
$\lcone \simeq \LG \thetaGo \simeq  \sqrt{ c\tauG\LG}$, \added{where $\tauG$ is the scattering time from the Milky Way, and $\LG$ is the distance between the observer and Galactic scattering screen}.   Requiring $\lc \gtrsim \lcone$ then yields the inequality,
\begin{equation}
    \tau_X \tau_G \lesssim \frac{1}{(2\pi\nu)^2}\frac{\dso^2}{L_X L_G} \approx (0.16 \ {\rm ms})^2 \times \frac{\dso^2}{\nu^2 L_X L_G}
\end{equation}
for $\dso$ in Gpc, $(L_X, L_G)$ in kpc, and $\nu$ in GHz \citep{2019ARA&A..57..417C}. For $\tau_G \approx 4$ $\mu$s (\added{where $\tau_G$ is related to the measured $\nud$ as $\nud \approx 1/2\pi\tauG$}; see Section~\ref{sec:nud_obs}) and $\dso \approx 810$ Mpc \added{\citep{2021arXiv211007418N}}, we find $L_X L_G \lesssim 0.1$ kpc$^2$. Owing to the FRB's high Galactic latitude, any Galactic scattering will be dominated by the thick disk within $L_G \sim 1$ kpc \citep{2020ApJ...897..124O}. Hence, the estimated distance between the extragalactic scattering screen and FRB source is $L_X \lesssim 0.1$ kpc, entirely consistent with scattering in the host galaxy. This upper limit is less than $2\%$ of the distance between the FRB source and peak star-forming region in the galaxy, and fourteen times smaller than the current upper limit on the size of the associated PRS, 1.4 kpc \citep{2021arXiv211007418N}. It therefore appears highly likely that the pulse broadening of FRB~190520 is dominated by gas within HG190520, including material near or within the FRB local environment. \added{Future improvements to VLBI constraints on the PRS size may reach milliarcsecond precision, which will probe scales well within the 100 pc region that appears relevant to scattering.}

\subsubsection{Host Galaxy Scattering Parameters}
The cloudlet model used in \S\ref{sec:jointDMHalpha} also yields an expression for the scattering time in terms of the DM contributed by a scattering layer $\DMl$ in $\DMunits$,
a source at redshift $z$, and a scattering layer at redshift $\zl$ \added{\citep{2021ApJ...911..102O}}, 
\begin{equation}
     \tau(\DM,\nu, z) \approx 48.03~{\rm \mu s} \times
       \frac{ \Atau  \FtG\,  \DMl^2 }{(1 + \zl)^3 \nu^4}, 
     	%\Gscatt.
\label{eq:taudmnuz}
\end{equation}
where the observing frequency $\nu$ is in GHz. 
The geometric factor $\Gscatt$ depends on the relative locations of the source, lens, and observer, and for scattering of a source located within a scattering medium at large distances from the observer, $\Gscatt = 1$. For scattering of a source at cosmological distances by an intervening galaxy, $\Gscatt = 2 \dsl \dlo/L\dso \gg 1$, where $L$ is the path length through the lens. 
The pre-factor in Equation~\ref{eq:taudmnuz} is for $L$ in Mpc and all other distances in Gpc. The dimensionless factor $\Atau$ converts the mean scattering delay to the $1/e$ time that is typically estimated from observed pulses. For the remainder of our analysis, we assume $\Atau\approx1$. The parameter $\Ftilde = \zeta\epsilon^2 / f (\louter^2\linner)^{1/3}$ \added{quantifies turbulent density fluctuations and} has units ${\rm (pc^2\ km)^{1/3}}$, where $\zeta$, $\epsilon^2$, and $f$ have been defined previously,
and $\louter$ and $\linner$ are the outer and inner scales of the turbulence wavenumber spectrum in pc and km, respectively 
\citep[][and references therein]{2021ApJ...911..102O}. \added{The product $\Ftilde G$ thus describes the combined amplification to scattering from both geometric effects and turbulent density fluctuations.}

\begin{figure}
    \centering
    \includegraphics[width=\linewidth]{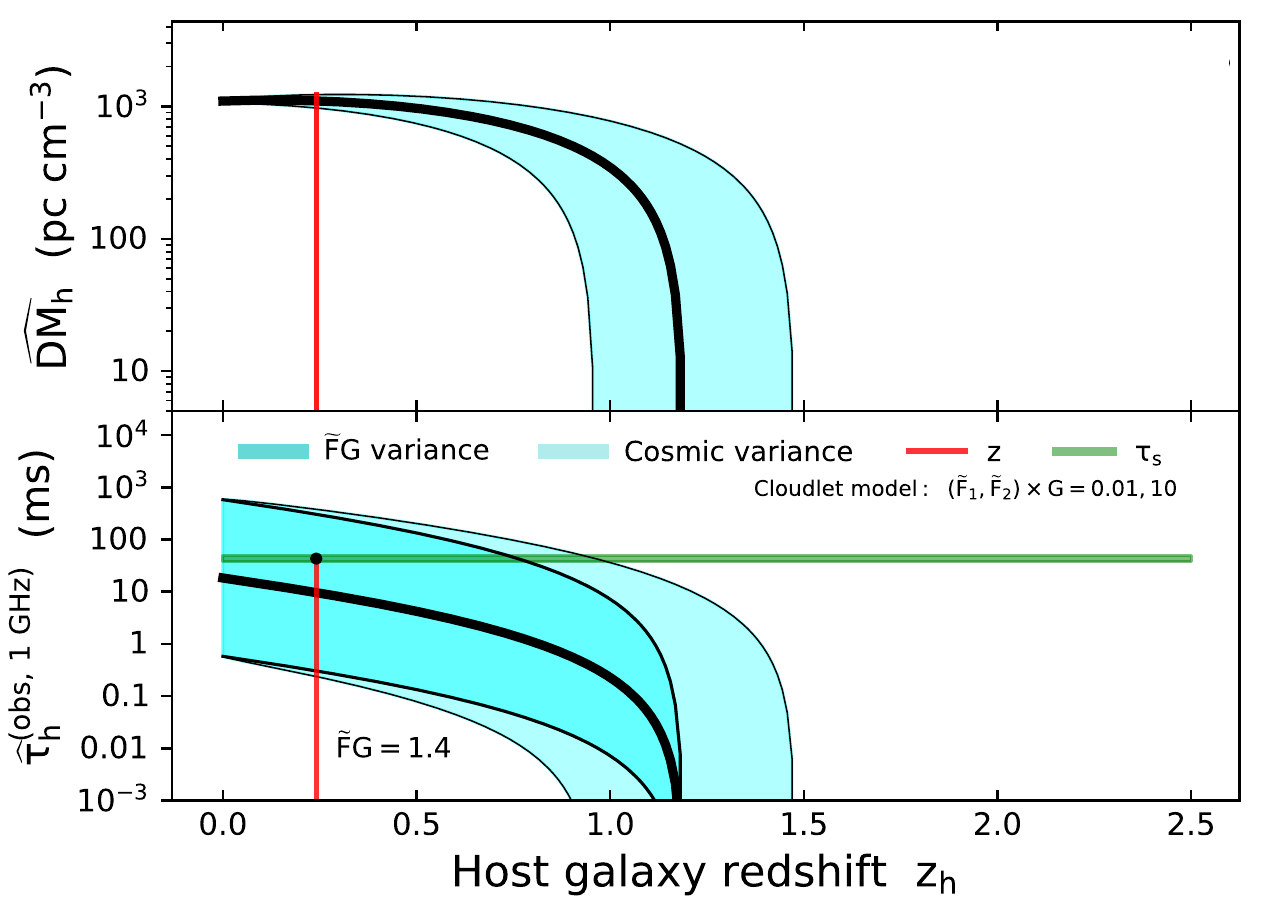}
    \caption{Constraints on dispersion measure and scattering time from the host galaxy vs redshift. The cyan shaded region in the top panel shows the range of values for $\DMh$ (in the rest frame of the host galaxy) taking into account uncertainties in the DM contributions from the Milky Way (disk + halo) and IGM, as described in the text.      The bottom panel shows the range of predicted scattering times $\tauhhat$ including cosmic variance in  $\DMigm$ (lighter cyan shading) and for a range of possible values for the combined parameter $\FtG$ from 0.1 to $10~\FtGunits$ (darker turqoise shading).
  The vertical red line indicates the measured redshift and the horizontal green line indicates the measured scattering time.  } 
    \label{fig:taudm_vs_z}
\end{figure}

\begin{figure}
    \centering
    \includegraphics[width=\linewidth]{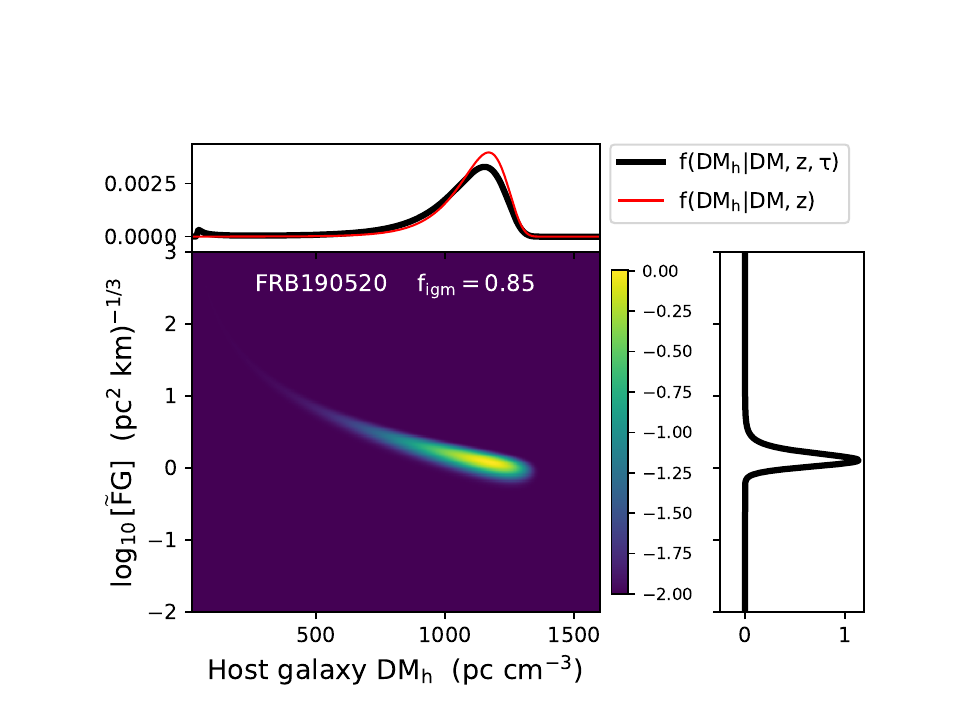}
    \caption{Posterior PDFs for $\DMh$ and $\FtG$ based on the DM inventory, measured scattering time $\tau$, and measured redshift.  The joint distribution is shown with a color bar indicating $\log_{10}$PDF.  The upper panel shows the  PDF of $\DMh$ (black curve) after marginalizing over $\FtG$. For comparison,  the red curve shows the posterior PDF using only the DM inventory and measured redshift.   The panel on the right is the PDF of $\FtG$ after marginalizing over $\DMh$. 
    }
    \label{fig:FG_vs_DMh}
\end{figure}

Figure~\ref{fig:taudm_vs_z} shows the combined estimates of the host galaxy's dispersion measure
$\DMhhat$  (top panel)  and scattering time estimate $\tauhhat$ (bottom panel)  vs. redshift, corresponding to the hypothetical case where the redshift is not known. The figure indicates that if the redshift had not been measured, a redshift up to nearly $\zh = 1.5$ would be allowed given uncertainties in the contributions from the Milky Way and IGM.  However, larger redshifts imply a smaller $\DMh$, making it less likely to account for the measured scattering time for a host galaxy ISM  similar to the Milky Way ISM.   If the measured DM were dominated by the IGM contribution, the scattering time would be highly anomalous for values of $\FtG$ encountered in the Milky Way.  Instead,  the scattering time is consistent with a large host DM using a reasonable value of $\FtG \sim 1.4~\FtGunits$ for the median value of $\DMh$. Figure~\ref{fig:FG_vs_DMh} shows the results of a full posterior analysis for $\DMh$ and $\FtG$, which yields a median value and 68\% probable region, 
$\FtG = 1.5^{+0.8}_{-0.3}~\FtGunits$.

% Keep: these are results from analyis of FtG quoted in the actual text
%FRB FRB190520 analysis from  post_analysis_DMh_FtG
%Ftilde G:
%    median =    1.523
%    median + =    2.299
%    median - =    1.189
    
\section{Redshift Estimation Using Dispersion Measures and Scattering Times}\label{sec:taudmz}

In \citet[][]{2021arXiv210801172C} we demonstrated that, in the absence of a direct measurement,  redshifts  estimated using a combined dispersion-scattering ($\DM$-$\tau$) estimator are less biased than those from  a
$\DM$-only based estimator and also have less scatter among the (only) nine FRB sources other than FRB 190520 with both redshift and scattering measurements available. Application of the same technique to FRB~190520 supports this conclusion more strongly:  a $\DM$-only based estimate yields a redshift that is too large by a factor of five
while incorporation of scattering brings the estimate in line with the measured redshift.   The bottom panel of
Figure~\ref{fig:taudm_vs_z} shows that the measured redshift designated by the vertical red line corresponds to a scattering  time estimated for a typical value of $\FtG$ that well matches the measured value (horizontal green line). 

\begin{figure}
    \centering
    \includegraphics[width=\linewidth]{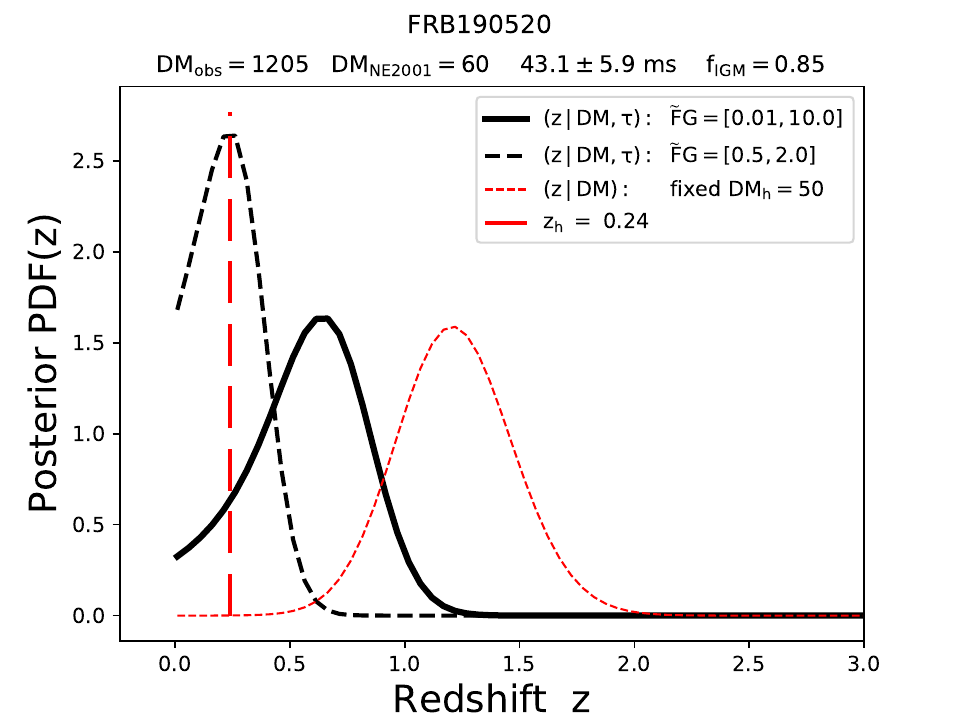}
    \caption{Posterior redshift PDFs for  FRB~190520 using three different redshift estimators based on dispersion and scattering  and using an IGM baryonic fraction $\figm = 0.85$.  Two (solid and dashed black lines) use  the measured DM along with the scattering time $\tau$ but with different ranges for $\FtG$ .  The third (thin red dotted line) uses only the measured DM.  The vertical, thick red dashed  line indicates the measured redshift of the associated host galaxy.
    }
    \label{fig:pdf_z}
\end{figure}

Figure~\ref{fig:pdf_z} shows  posterior PDFs for the redshift
of FRB~190520 using three redshift estimators that are independent of the measured redshift \citep[][]{2021arXiv210801172C}:

\begin{enumerate}
\item A DM-based estimate, $\zhat(\DM \vert \DMh)$, that uses only the DM inventory and a fixed contribution from  the host galaxy, $\DMh = 50~\DMunits$. 

\item A combined DM and scattering-based estimate, 
$\zhat(\DM, \tau \vert \FtG \in [0.5, 2]$),  that 
(statistically) matches the scattering time calculated from Eq.~\ref{eq:taudmnuz} with the  measured  scattering time to constrain $\DMh$ jointly with the 
 DM inventory.  For this case, Eq.~\ref{eq:taudmnuz} is employed  using a narrow range of $\FtG$ from 0.5 to $2~\FtGunits$.  
 
\item A second DM-scattering estimator, \replaced{$\zhat(\DM, \tau \vert \FtG = [0.1, 10])$}{$\zhat(\DM, \tau \vert \FtG = [0.01, 10])$},  that  uses a wider 
range of $\FtG$ values. 
\end{enumerate}

The cases shown are based on $\figm = 0.85$, a value that minimizes bias and scatter of redshift estimates for the FRBs analyzed in \citet[][]{2021arXiv210801172C}.  For FRB~190520 the best case is the DM-scattering estimator using a narrow range of $\FtG$, although use of the wider range is also consistent with the measured redshift with reasonable probability. However, the DM-only estimator is, not surprisingly, highly inconsistent.\added{In general the preferred range of $\FtG$ is LOS-dependent, and FRB~190520 is coincidentally consistent with a narrower range of $\FtG$ than other FRBs considered in \citet[][]{2021arXiv210801172C}.} 

\begin{figure}
    \centering
    \includegraphics[width=\linewidth]{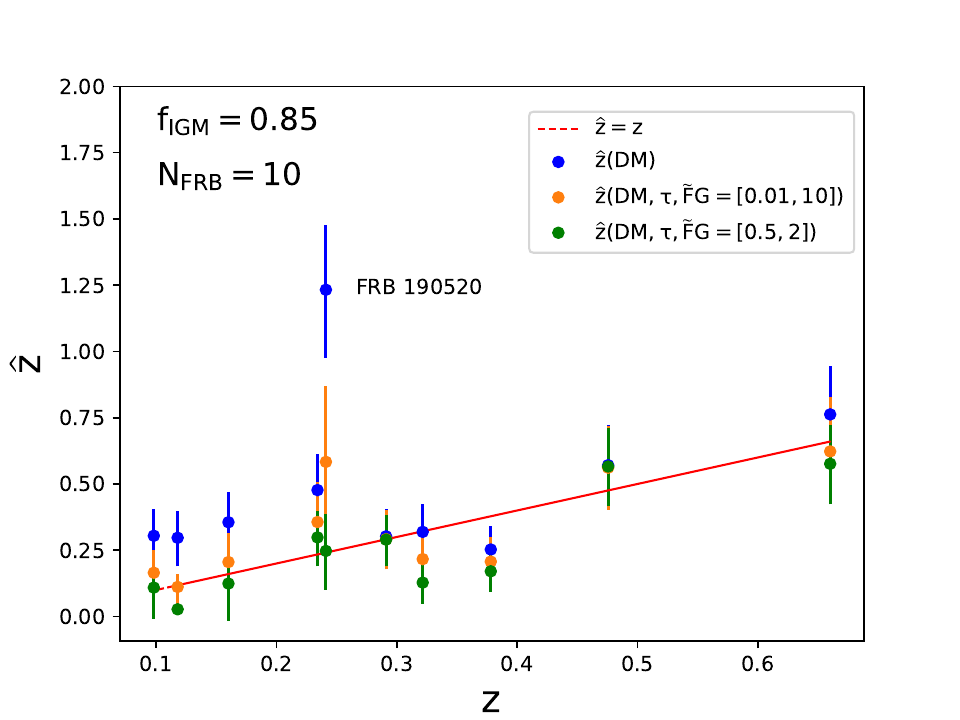}
    \caption{Estimated redshift $\zhat$ vs observed redshift $z$ using three different redshift estimators, as indicated in the legend, applied to ten FRBs that have both redshift and scattering time measurements. The slanted red line designates $\zhat = z$. 
    }
    \label{fig:zhat_vs_z_three}
\end{figure}

The performance of these estimators on FRB~190520 is compared in Figure~\ref{fig:zhat_vs_z_three} with the nine other FRBs analyzed previously that had both scattering and redshift measurements (these include: FRBs 180924, 181112, 190102, 190523, 190608, 190611, 191001, 200430, and 20201124A). Error bars on redshift estimates represent 68\% probable regions centered on median redshift values calculated from the posterior PDFs. Four of the FRBs shown in Figure~\ref{fig:zhat_vs_z_three}, 181112, 190523, 191001, and 200430 have both scattering in their host galaxies and  values of $\DMh > 200~\DMunits$ (comparable to that of FRB 121102), and FRB 200430 has been proposed as a promising candidate for PRS searches \citep{2022ApJ...927...55L}.

Inclusion of FRB~190520 in the sample yields the same overall result as in \citet[][]{2021arXiv210801172C} that inclusion of scattering dramatically reduces the scatter of redshift measurements.  In addition, this methodology allows a preferred range of $\figm$ to be identified, as demonstrated and discussed in \citet[][]{2021arXiv210801172C}. The value of $\FtG$ inferred for FRB 190520 lies within the range found for the other FRBs examined in \cite{2021arXiv210801172C} and shown in Figure~\ref{fig:zhat_vs_z_three}; this range of $\FtG$ is consistent with values inferred for the Milky Way ISM using Galactic pulsars. \added{Larger samples of localized FRBs will reveal under what conditions $\FtG$ in other galaxies differs from values found in the Milky Way and whether/how $\FtG$ varies between different galaxies as a function of redshift, which will in turn improve calibration of $\FtG$ for redshift estimation of non-localized sources.}

\section{Discussion and Summary}\label{sec:discussion}

\indent The host galaxy of FRB 190520 exhibits extreme plasma properties compared to the host galaxies of other localized FRBs. The substantially lower than expected redshift implies that the host galaxy  dominates the FRB DM budget. Given the small DM contribution expected from the host galaxy halo, the majority of the host DM likely originates in the host ISM and the FRB circum-source environment. A large $\DMh$ is found through both the DM inventory, which yields $\DMh = 1121^{+89}_{-138}\ \DMunits$ (host frame), and through Balmer line observations. Assuming a nominal temperature of $T\sim10^4$ K for the H$\alpha$ emitting gas yields a DM of about 300 \DMunit\ (host frame), which is still large compared to the DMs of other FRB host galaxies, but is significantly lower than the $\DMh$ inferred from the FRB DM inventory. The H$\alpha$ emission could be made consistent with a much larger DM if the gas temperature and density fluctuation statistics are significantly different from those considered typical of the warm ionized medium in the Milky Way. 
The FRB DM also receives contributions from the IGM and Milky Way, but these contributions comprise less than $20\%$ of the total DM budget for this LOS.

\indent Negligible scattering is expected from the IGM at the FRB redshift \citep{2013ApJ...776..125M}. We constrain the scattering contributions from the host galaxy and Milky Way along this LOS through measurements of scatter broadening and scintillation. The FRB mean scattering time of $10.9\pm1.5$ ms at 1.41 GHz, equivalent to $\tau \sim 300$ ms at 0.6 GHz, is larger than any of the scattering times observed in CHIME/FRB Catalog 1, which only contains two FRBs with $\tau > 50$ ms \citep{2021ApJS..257...59A}. Correcting for selection biases in CHIME/FRB Catalog 1 suggests there should be a substantial population of highly scattered FRBs \citep{2021ApJS..257...59A}, of which FRB 190520 is clearly an example. 

\indent Combining $\tau$ with $\DMh$ for FRB 190520 yields a value for $\FtG=1.5^{+0.8}_{-0.3}\ \FtGunits$ that is similar to values of $\FtG$ found for other FRBs with both measured redshifts and scattering attributable to their host galaxies \citep[see Section~\ref{sec:taudmz}, and][]{2021arXiv210801172C}. This result affirms that a scattering-based redshift estimator can produce more robust redshift predictions than a DM-only redshift estimator, when scattering is observed from the host galaxy. Combined with the mean scintillation bandwidth $\nud = 0.21\pm0.01$ MHz at 1.41 GHz, the measured scattering time implies a distance between the FRB source and dominant extragalactic scattering layer $L_X \lesssim 100$ pc. This upper bound on $L_X$ is far more stringent than upper limits inferred from the scintillation bandwidths of other localized FRBs \citep[e.g.][]{2015Natur.528..523M}, and could suggest that scattering occurs in the FRB circum-source environment. %\textbf{If FRB 190520 is embedded in a synchrotron nebula similar to that of FRB 121102, then the distance to the circum-source scattering layer may be as small as $\sim 10^{-2} - 10^{-1}$ pc (based on a lack of synchrotron self-absorption observed from the FRB 121102 PRS; see \citealt{2018ApJ...868L...4M}). Further analysis of the PRS spectrum for FRB 190520 may yield independent constraints on the size of the FRB source environment.}

\indent If FRB 190520 is embedded in a synchrotron-emitting nebula similar to that proposed for FRB 121102, then the lack of a synchrotron self-absorption signature in the observed PRS spectrum down to 1.4 GHz \citep{2021arXiv211007418N} yields an approximate lower limit on the size of the nebula $R_n$. Using the formalism of \cite{2018ApJ...868L...4M} and the observed PRS flux density at 3 GHz $\approx 200$ $\mu$Jy, we find $R_n \lesssim 0.9\times10^{17}$ cm $\approx0.03$ pc, over three orders of magnitude smaller than the upper limit on $L_X$. This constraint on $R_n$ is very similar to that of the FRB 121102 PRS, due to their comparable flux densities and distances. The relativistic electrons responsible for the PRS would not contribute to dispersion or scattering, but the upper limit on $L_X$ could potentially support a scenario where dispersion and scattering arise within a supernova remnant or merger ejecta surrounding a magnetar and synchrotron nebula \citep{2018MNRAS.481.2407M}. \added{This physical model may also be relevant to the extreme RM variations observed from FRB 190520, which may originate within the plasma region that also appears relevant to scattering \citep{Fengscience,2022arXiv220211112A,2022arXiv220308151D}.}

\indent Further disentangling the host galaxy ISM from the FRB near-source environment in terms of their DM and scattering contributions involves several factors that are not well-constrained. The only independent constraint on the host galaxy DM comes from the observed H$\alpha$ emission, but it is unclear how much of that H$\alpha$ emission is traced by the FRB LOS. It is also unclear whether the entire measured scattering time could be attributed to the FRB circum-source environment. Moving the scattering plasma layer closer to the source reduces the geometric leverage to scattering, quantified as $s(1-s/D)$, where $s$ is the fractional screen location ($s=0$ at the source and $1$ at the observer) and $D$ is the total distance between the source and observer. For a screen very close to the source, $s \ll 1$, and a corresponding increase in the level of turbulence (quantified as $\cnsq$ or slightly differently as $\Ftilde$) is required to produce the same amount of scattering. 

\indent A smaller distance between the source and scattering layer also reduces the allowed width of the scattering layer. If the entire observed DM comes from this same scattering layer, then the layer's mean electron density $n_e$ must also increase. For a plasma layer with $\DM\sim1000\ \DMunits$ and a width of order $L_X\sim 100$ pc, $n_e \sim 10$ cm$^{-3}$, which implies an $\rm EM\sim 10^4$ pc cm$^{-6}$ and a negligible optical depth due to free-free absorption at 1.4 GHz. However, if the layer width decreases to $1$ pc then $n_e\sim1000$ cm$^{-3}$ and $\rm EM\sim10^8$ pc cm$^{-6}$, and reducing the width to $0.1$ pc yields $n_e\sim10^4$ cm$^{-3}$ and $\rm EM\sim10^{12}$ pc cm$^{-6}$. At such high densities free-free absorption can play a role in FRB detectability, although it has been demonstrated that free-free absorption can be suppressed in plasma within $\sim1$ pc of FRB sources due to Coulomb collisions in the surrounding plasma \citep{2020MNRAS.496.3308L}. We have assumed here that the scattering and DM are contributed by a single plasma layer, but a more complex plasma configuration within the host galaxy is entirely possible. 

\indent Observations of DM and scattering are sensitive to path-integrated electron column densities and fluctuations. While complementary observations of scatter broadening, Galactic DISS, and angular broadening can be used to infer where scattering occurs along an FRB LOS and even within a host galaxy, deconstructing the DM budget within a host galaxy to infer properties of the host ISM and FRB near-source environment will benefit from higher spatial resolution H$\alpha$ measurements and observations at complementary wavelengths \citep[e.g.][]{2021ApJ...922..173C,2021ApJ...908L..12T}. It is still unclear whether the unusually large DM and scattering observed from FRB 190520 constitute unique features of its near-source environment and/or its host galaxy ISM more broadly; however, given both the large H$\alpha$ EM and the stringent upper limit on the distance between the FRB source and dominant scattering plasma, it appears likely that both regions (ISM and circum-source) contribute significantly to the total DM and scattering observed. Regardless, our results are consistent with previous findings that the scattering of localized FRBs can be accounted for by plasma in their host galaxies and the Milky Way. Continued application of detailed DM and scattering budgets will not only resolve the plasma density and turbulence within the distant galactic environments of localized FRBs \citep[e.g.][]{2021arXiv210711334S}, but will also improve redshift estimates for FRBs that have not yet been associated with their host galaxies. \added{Such improvements will in turn inform the use of FRBs as cosmological probes, including constraints on the IGM and other galaxies along FRB LOSs.}

\acknowledgements{The authors thank the anonymous referee and R. Main for comments that improved this work. SKO, JMC, and SC acknowledge support from the National Science Foundation (AAG~1815242) and are members of the NANOGrav Physics Frontiers Center, which is supported by NSF award PHY-2020265. CHN is supported by the FAST Fellowship and DL acknowledges support from the National Natural Science Foundation of China
(NSFC) Programs No. 11988101 and No. 11725313. CWT acknowledges support from NSFC No. 11973051. JWM is a CITA Postdoctoral Fellow supported by the Natural Sciences and Engineering Research Council of Canada (NSERC), [funding reference \#CITA 490888-16]. CJL acknowledges support from the National Science Foundation under Grant No.~2022546. RAT acknowledges support from NSF grant AAG-1714897.} 

\bibliography{bib,jmcadds}

\begin{thebibliography}{}
\expandafter\ifx\csname natexlab\endcsname\relax\def\natexlab#1{#1}\fi
\providecommand{\url}[1]{\href{#1}{#1}}
\providecommand{\dodoi}[1]{doi:~\href{http://doi.org/#1}{\nolinkurl{#1}}}
\providecommand{\doeprint}[1]{\href{http://ascl.net/#1}{\nolinkurl{http://ascl.net/#1}}}
\providecommand{\doarXiv}[1]{\href{https://arxiv.org/abs/#1}{\nolinkurl{https://arxiv.org/abs/#1}}}

\bibitem[{{Amiri} {et~al.}(2021){Amiri}, {Andersen}, {Bandura}, {Berger},
  {Bhardwaj}, {Boyce}, {Boyle}, {Brar}, {Breitman}, {Cassanelli}, {Chawla},
  {Chen}, {Cliche}, {Cook}, {Cubranic}, {Curtin}, {Deng}, {Dobbs}, {(Adam)
  Dong}, {Eadie}, {Fandino}, {Fonseca}, {Gaensler}, {Giri}, {Good}, {Halpern},
  {Hill}, {Hinshaw}, {Josephy}, {Kaczmarek}, {Kader}, {Kania}, {Kaspi},
  {Landecker}, {Lang}, {Leung}, {Li}, {Lin}, {Masui}, {McKinven}, {Mena-Parra},
  {Merryfield}, {Meyers}, {Michilli}, {Milutinovic}, {Mirhosseini},
  {M{\"u}nchmeyer}, {Naidu}, {Newburgh}, {Ng}, {Patel}, {Pen}, {Petroff},
  {Pinsonneault-Marotte}, {Pleunis}, {Rafiei-Ravandi}, {Rahman}, {Ransom},
  {Renard}, {Sanghavi}, {Scholz}, {Shaw}, {Shin}, {Siegel}, {Sikora}, {Singh},
  {Smith}, {Stairs}, {Tan}, {Tendulkar}, {Vanderlinde}, {Wang}, {Wulf},
  {Zwaniga}, \& {CHIME/FRB Collaboration}}]{2021ApJS..257...59A}
{Amiri}, M., {Andersen}, B.~C., {Bandura}, K., {et~al.} 2021, ApJS, 257, 59,
  \dodoi{10.3847/1538-4365/ac33ab}

\bibitem[{{Anna-Thomas} {et~al.}(2022){Anna-Thomas}, {Connor}, {Burke-Spolaor},
  {Beniamini}, {Aggarwal}, {Law}, {Lynch}, {Li}, {Feng}, {Ocker}, {Cruces},
  {Chatterjee}, {Yu}, {Niu}, \& {Xue}}]{2022arXiv220211112A}
{Anna-Thomas}, R., {Connor}, L., {Burke-Spolaor}, S., {et~al.} 2022, arXiv
  e-prints, arXiv:2202.11112.
\newblock \doarXiv{2202.11112}

\bibitem[{{Astropy Collaboration} {et~al.}(2013){Astropy Collaboration},
  {Robitaille}, {Tollerud}, {Greenfield}, {Droettboom}, {Bray}, {Aldcroft},
  {Davis}, {Ginsburg}, {Price-Whelan}, {Kerzendorf}, {Conley}, {Crighton},
  {Barbary}, {Muna}, {Ferguson}, {Grollier}, {Parikh}, {Nair}, {Unther},
  {Deil}, {Woillez}, {Conseil}, {Kramer}, {Turner}, {Singer}, {Fox}, {Weaver},
  {Zabalza}, {Edwards}, {Azalee Bostroem}, {Burke}, {Casey}, {Crawford},
  {Dencheva}, {Ely}, {Jenness}, {Labrie}, {Lim}, {Pierfederici}, {Pontzen},
  {Ptak}, {Refsdal}, {Servillat}, \& {Streicher}}]{2013AA...558A..33A}
{Astropy Collaboration}, {Robitaille}, T.~P., {Tollerud}, E.~J., {et~al.} 2013,
  A\&A, 558, A33, \dodoi{10.1051/0004-6361/201322068}

\bibitem[{{Astropy Collaboration} {et~al.}(2018){Astropy Collaboration},
  {Price-Whelan}, {Sip{\H{o}}cz}, {G{\"u}nther}, {Lim}, {Crawford}, {Conseil},
  {Shupe}, {Craig}, {Dencheva}, {Ginsburg}, {VanderPlas}, {Bradley},
  {P{\'e}rez-Su{\'a}rez}, {de Val-Borro}, {Aldcroft}, {Cruz}, {Robitaille},
  {Tollerud}, {Ardelean}, {Babej}, {Bach}, {Bachetti}, {Bakanov}, {Bamford},
  {Barentsen}, {Barmby}, {Baumbach}, {Berry}, {Biscani}, {Boquien}, {Bostroem},
  {Bouma}, {Brammer}, {Bray}, {Breytenbach}, {Buddelmeijer}, {Burke},
  {Calderone}, {Cano Rodr{\'\i}guez}, {Cara}, {Cardoso}, {Cheedella}, {Copin},
  {Corrales}, {Crichton}, {D'Avella}, {Deil}, {Depagne}, {Dietrich}, {Donath},
  {Droettboom}, {Earl}, {Erben}, {Fabbro}, {Ferreira}, {Finethy}, {Fox},
  {Garrison}, {Gibbons}, {Goldstein}, {Gommers}, {Greco}, {Greenfield},
  {Groener}, {Grollier}, {Hagen}, {Hirst}, {Homeier}, {Horton}, {Hosseinzadeh},
  {Hu}, {Hunkeler}, {Ivezi{\'c}}, {Jain}, {Jenness}, {Kanarek}, {Kendrew},
  {Kern}, {Kerzendorf}, {Khvalko}, {King}, {Kirkby}, {Kulkarni}, {Kumar},
  {Lee}, {Lenz}, {Littlefair}, {Ma}, {Macleod}, {Mastropietro}, {McCully},
  {Montagnac}, {Morris}, {Mueller}, {Mumford}, {Muna}, {Murphy}, {Nelson},
  {Nguyen}, {Ninan}, {N{\"o}the}, {Ogaz}, {Oh}, {Parejko}, {Parley}, {Pascual},
  {Patil}, {Patil}, {Plunkett}, {Prochaska}, {Rastogi}, {Reddy Janga},
  {Sabater}, {Sakurikar}, {Seifert}, {Sherbert}, {Sherwood-Taylor}, {Shih},
  {Sick}, {Silbiger}, {Singanamalla}, {Singer}, {Sladen}, {Sooley},
  {Sornarajah}, {Streicher}, {Teuben}, {Thomas}, {Tremblay}, {Turner},
  {Terr{\'o}n}, {van Kerkwijk}, {de la Vega}, {Watkins}, {Weaver}, {Whitmore},
  {Woillez}, {Zabalza}, \& {Astropy Contributors}}]{2018AJ....156..123A}
{Astropy Collaboration}, {Price-Whelan}, A.~M., {Sip{\H{o}}cz}, B.~M., {et~al.}
  2018, AJ, 156, 123, \dodoi{10.3847/1538-3881/aabc4f}

\bibitem[{{Bij} {et~al.}(2021){Bij}, {Lin}, {Li}, {van Kerkwijk}, {Pen}, {Lu},
  {Main}, {Peterson}, {Quine}, \& {Vanderlinde}}]{2021ApJ...920...38B}
{Bij}, A., {Lin}, H.-H., {Li}, D., {et~al.} 2021, ApJ, 920, 38,
  \dodoi{10.3847/1538-4357/ac1589}

\bibitem[{{Bochenek} {et~al.}(2020){Bochenek}, {Ravi}, {Belov}, {Hallinan},
  {Kocz}, {Kulkarni}, \& {McKenna}}]{brb+20}
{Bochenek}, C.~D., {Ravi}, V., {Belov}, K.~V., {et~al.} 2020, Nature, 587, 59,
  \dodoi{10.1038/s41586-020-2872-x}

\bibitem[{{Caleb} {et~al.}(2022){Caleb}, {Rajwade}, {Desvignes}, {Stappers},
  {Lyne}, {Weltevrede}, {Kramer}, {Levin}, \& {Surnis}}]{crd+22}
{Caleb}, M., {Rajwade}, K., {Desvignes}, G., {et~al.} 2022, MNRAS, 510, 1996,
  \dodoi{10.1093/mnras/stab3223}

\bibitem[{{Chittidi} {et~al.}(2021){Chittidi}, {Simha}, {Mannings},
  {Prochaska}, {Ryder}, {Rafelski}, {Neeleman}, {Macquart}, {Tejos},
  {Jorgenson}, {Day}, {Marnoch}, {Bhandari}, {Deller}, {Qiu}, {Bannister},
  {Shannon}, \& {Heintz}}]{2021ApJ...922..173C}
{Chittidi}, J.~S., {Simha}, S., {Mannings}, A., {et~al.} 2021, ApJ, 922, 173,
  \dodoi{10.3847/1538-4357/ac2818}

\bibitem[{{Cordes} \& {Chatterjee}(2019)}]{2019ARA&A..57..417C}
{Cordes}, J.~M., \& {Chatterjee}, S. 2019, ARAA, 57, 417,
  \dodoi{10.1146/annurev-astro-091918-104501}

\bibitem[{{Cordes} \& {Lazio}(2002)}]{2002astro.ph..7156C}
{Cordes}, J.~M., \& {Lazio}, T.~J.~W. 2002, arXiv e-prints, astro.
\newblock \doarXiv{astro-ph/0207156}

\bibitem[{{Cordes} {et~al.}(2021){Cordes}, {Ocker}, \&
  {Chatterjee}}]{2021arXiv210801172C}
{Cordes}, J.~M., {Ocker}, S.~K., \& {Chatterjee}, S. 2021, arXiv e-prints,
  arXiv:2108.01172.
\newblock \doarXiv{2108.01172}

\bibitem[{{Cordes} \& {Wasserman}(2016)}]{2016MNRAS.457..232C}
{Cordes}, J.~M., \& {Wasserman}, I. 2016, MNRAS, 457, 232,
  \dodoi{10.1093/mnras/stv2948}

\bibitem[{{Cordes} {et~al.}(2017){Cordes}, {Wasserman}, {Hessels}, {Lazio},
  {Chatterjee}, \& {Wharton}}]{2017ApJ...842...35C}
{Cordes}, J.~M., {Wasserman}, I., {Hessels}, J.~W.~T., {et~al.} 2017, ApJ, 842,
  35, \dodoi{10.3847/1538-4357/aa74da}

\bibitem[{{Cordes} {et~al.}(1991){Cordes}, {Weisberg}, {Frail}, {Spangler}, \&
  {Ryan}}]{1991Natur.354..121C}
{Cordes}, J.~M., {Weisberg}, J.~M., {Frail}, D.~A., {Spangler}, S.~R., \&
  {Ryan}, M. 1991, Nature, 354, 121, \dodoi{10.1038/354121a0}

\bibitem[{{Dai} {et~al.}(2022){Dai}, {Feng}, {Yang}, {Zhang}, {Li}, {Niu},
  {Wang}, {Xue}, {Zhang}, {Burke-Spolaor}, {Law}, {Lynch}, {Connor},
  {Anna-Thomas}, {Zhang}, {Duan}, {Yao}, {Tsai}, {Zhu}, {Cruces}, {Hobbs},
  {Miao}, {Niu}, {Filipovic}, \& {Zhu}}]{2022arXiv220308151D}
{Dai}, S., {Feng}, Y., {Yang}, Y.~P., {et~al.} 2022, arXiv e-prints,
  arXiv:2203.08151.
\newblock \doarXiv{2203.08151}

\bibitem[{{Draine}(2011)}]{2011piim.book.....D}
{Draine}, B.~T. 2011, {Physics of the Interstellar and Intergalactic Medium}
  (Princeton University Press)

\bibitem[{Feng {et~al.}(2022)Feng, Li, Yang, Zhang, Zhu, Zhang, Lu, Wang, Dai,
  Lynch, Yao, Jiang, Niu, Zhou, Xu, Miao, Niu, Meng, Qian, Tsai, Wang, Xue,
  Yue, Yuan, Zhang, \& Zhang}]{Fengscience}
Feng, Y., Li, D., Yang, Y.-P., {et~al.} 2022, Science, 375, 1266,
  \dodoi{10.1126/science.abl7759}

\bibitem[{{Gwinn} {et~al.}(1998){Gwinn}, {Britton}, {Reynolds}, {Jauncey},
  {King}, {McCulloch}, {Lovell}, \& {Preston}}]{1998ApJ...505..928G}
{Gwinn}, C.~R., {Britton}, M.~C., {Reynolds}, J.~E., {et~al.} 1998, ApJ, 505,
  928, \dodoi{10.1086/306178}

\bibitem[{{Harvey-Smith} {et~al.}(2011){Harvey-Smith}, {Madsen}, \&
  {Gaensler}}]{2011ApJ...736...83H}
{Harvey-Smith}, L., {Madsen}, G.~J., \& {Gaensler}, B.~M. 2011, ApJ, 736, 83,
  \dodoi{10.1088/0004-637X/736/2/83}

\bibitem[{{Hessels} {et~al.}(2019){Hessels}, {Spitler}, {Seymour}, {Cordes},
  {Michilli}, {Lynch}, {Gourdji}, {Archibald}, {Bassa}, {Bower}, {Chatterjee},
  {Connor}, {Crawford}, {Deneva}, {Gajjar}, {Kaspi}, {Keimpema}, {Law},
  {Marcote}, {McLaughlin}, {Paragi}, {Petroff}, {Ransom}, {Scholz}, {Stappers},
  \& {Tendulkar}}]{2019ApJ...876L..23H}
{Hessels}, J.~W.~T., {Spitler}, L.~G., {Seymour}, A.~D., {et~al.} 2019, ApJL,
  876, L23, \dodoi{10.3847/2041-8213/ab13ae}

\bibitem[{{Law} {et~al.}(2022){Law}, {Connor}, \&
  {Aggarwal}}]{2022ApJ...927...55L}
{Law}, C.~J., {Connor}, L., \& {Aggarwal}, K. 2022, ApJ, 927, 55,
  \dodoi{10.3847/1538-4357/ac4c42}

\bibitem[{{Li} {et~al.}(2019){Li}, {Dickey}, \& {Liu}}]{li19}
{Li}, D., {Dickey}, J.~M., \& {Liu}, S. 2019, Research in Astronomy and
  Astrophysics, 19, 016, \dodoi{10.1088/1674-4527/19/2/16}

\bibitem[{{Li} {et~al.}(2018){Li}, {Wang}, {Qian}, {Krco}, {Jiang}, {Yue},
  {Jin}, {Zhu}, {Pan}, {Nan}, \& {Dunning}}]{li18}
{Li}, D., {Wang}, P., {Qian}, L., {et~al.} 2018, IEEE Microwave Magazine, 19,
  112, \dodoi{10.1109/MMM.2018.2802178}

\bibitem[{{Lu} \& {Phinney}(2020)}]{2020MNRAS.496.3308L}
{Lu}, W., \& {Phinney}, E.~S. 2020, MNRAS, 496, 3308,
  \dodoi{10.1093/mnras/staa1679}

\bibitem[{{Lyutikov}(2021)}]{lyu21}
{Lyutikov}, M. 2021, ApJ, 922, 166, \dodoi{10.3847/1538-4357/ac1b32}

\bibitem[{{Macquart} \& {Koay}(2013)}]{2013ApJ...776..125M}
{Macquart}, J.-P., \& {Koay}, J.~Y. 2013, ApJ, 776, 125,
  \dodoi{10.1088/0004-637X/776/2/125}

\bibitem[{{Macquart} {et~al.}(2020){Macquart}, {Prochaska}, {McQuinn},
  {Bannister}, {Bhandari}, {Day}, {Deller}, {Ekers}, {James}, {Marnoch},
  {Os{\l}owski}, {Phillips}, {Ryder}, {Scott}, {Shannon}, \&
  {Tejos}}]{2020Natur.581..391M}
{Macquart}, J.~P., {Prochaska}, J.~X., {McQuinn}, M., {et~al.} 2020, Nature,
  581, 391, \dodoi{10.1038/s41586-020-2300-2}

\bibitem[{{Main} {et~al.}(2022){Main}, {Hilmarsson}, {Marthi}, {Spitler},
  {Wharton}, {Bethapudi}, {Li}, \& {Lin}}]{2022MNRAS.509.3172M}
{Main}, R.~A., {Hilmarsson}, G.~H., {Marthi}, V.~R., {et~al.} 2022, MNRAS, 509,
  3172, \dodoi{10.1093/mnras/stab3218}

\bibitem[{{Marcote} {et~al.}(2020){Marcote}, {Nimmo}, {Hessels}, {Tendulkar},
  {Bassa}, {Paragi}, {Keimpema}, {Bhardwaj}, {Karuppusamy}, {Kaspi}, {Law},
  {Michilli}, {Aggarwal}, {Andersen}, {Archibald}, {Bandura}, {Bower}, {Boyle},
  {Brar}, {Burke-Spolaor}, {Butler}, {Cassanelli}, {Chawla}, {Demorest},
  {Dobbs}, {Fonseca}, {Giri}, {Good}, {Gourdji}, {Josephy}, {Kirichenko},
  {Kirsten}, {Landecker}, {Lang}, {Lazio}, {Li}, {Lin}, {Linford}, {Masui},
  {Mena-Parra}, {Naidu}, {Ng}, {Patel}, {Pen}, {Pleunis}, {Rafiei-Ravandi},
  {Rahman}, {Renard}, {Scholz}, {Siegel}, {Smith}, {Stairs}, {Vanderlinde}, \&
  {Zwaniga}}]{2020Natur.577..190M}
{Marcote}, B., {Nimmo}, K., {Hessels}, J.~W.~T., {et~al.} 2020, Nature, 577,
  190, \dodoi{10.1038/s41586-019-1866-z}

\bibitem[{{Margalit} \& {Metzger}(2018)}]{2018ApJ...868L...4M}
{Margalit}, B., \& {Metzger}, B.~D. 2018, ApJL, 868, L4,
  \dodoi{10.3847/2041-8213/aaedad}

\bibitem[{{Margalit} {et~al.}(2018){Margalit}, {Metzger}, {Berger}, {Nicholl},
  {Eftekhari}, \& {Margutti}}]{2018MNRAS.481.2407M}
{Margalit}, B., {Metzger}, B.~D., {Berger}, E., {et~al.} 2018, MNRAS, 481,
  2407, \dodoi{10.1093/mnras/sty2417}

\bibitem[{{Masui} {et~al.}(2015){Masui}, {Lin}, {Sievers}, {Anderson}, {Chang},
  {Chen}, {Ganguly}, {Jarvis}, {Kuo}, {Li}, {Liao}, {McLaughlin}, {Pen},
  {Peterson}, {Roman}, {Timbie}, {Voytek}, \& {Yadav}}]{2015Natur.528..523M}
{Masui}, K., {Lin}, H.-H., {Sievers}, J., {et~al.} 2015, Nature, 528, 523,
  \dodoi{10.1038/nature15769}

\bibitem[{{Nan} {et~al.}(2011){Nan}, {Li}, {Jin}, {Wang}, {Zhu}, {Zhu},
  {Zhang}, {Yue}, \& {Qian}}]{nan11}
{Nan}, R., {Li}, D., {Jin}, C., {et~al.} 2011, International Journal of Modern
  Physics D, 20, 989, \dodoi{10.1142/S0218271811019335}

\bibitem[{{Nimmo} {et~al.}(2022){Nimmo}, {Hessels}, {Kirsten}, {Keimpema},
  {Cordes}, {Snelders}, {Hewitt}, {Karuppusamy}, {Archibald}, {Bezrukovs},
  {Bhardwaj}, {Blaauw}, {Buttaccio}, {Cassanelli}, {Conway}, {Corongiu},
  {Feiler}, {Fonseca}, {Forss{\'e}n}, {Gawro{\'n}ski}, {Giroletti}, {Kharinov},
  {Leung}, {Lindqvist}, {Maccaferri}, {Marcote}, {Masui}, {Mckinven},
  {Melnikov}, {Michilli}, {Mikhailov}, {Ng}, {Orbidans}, {Ould-Boukattine},
  {Paragi}, {Pearlman}, {Petroff}, {Rahman}, {Scholz}, {Shin}, {Smith},
  {Stairs}, {Surcis}, {Tendulkar}, {Vlemmings}, {Wang}, {Yang}, \&
  {Yuan}}]{2022NatAs...6..393N}
{Nimmo}, K., {Hessels}, J.~W.~T., {Kirsten}, F., {et~al.} 2022, Nature
  Astronomy, 6, 393, \dodoi{10.1038/s41550-021-01569-9}

\bibitem[{{Niu} {et~al.}(2021){Niu}, {Aggarwal}, {Li}, {Zhang}, {Chatterjee},
  {Tsai}, {Yu}, {Law}, {Burke-Spolaor}, {Cordes}, {Zhang}, {Ocker}, {Yao},
  {Wang}, {Feng}, {Niino}, {Bochenek}, {Cruces}, {Connor}, {Jiang}, {Dai},
  {Luo}, {Li}, {Miao}, {Niu}, {Anna-Thomas}, {Sydnor}, {Stern}, {Wang}, {Yuan},
  {Yue}, {Zhou}, {Yan}, {Zhu}, \& {Zhang}}]{2021arXiv211007418N}
{Niu}, C.~H., {Aggarwal}, K., {Li}, D., {et~al.} 2021, arXiv e-prints,
  arXiv:2110.07418v2.
\newblock \doarXiv{2110.07418v2}

\bibitem[{{Ocker} {et~al.}(2020){Ocker}, {Cordes}, \&
  {Chatterjee}}]{2020ApJ...897..124O}
{Ocker}, S.~K., {Cordes}, J.~M., \& {Chatterjee}, S. 2020, ApJ, 897, 124,
  \dodoi{10.3847/1538-4357/ab98f9}

\bibitem[{{Ocker} {et~al.}(2021){Ocker}, {Cordes}, \&
  {Chatterjee}}]{2021ApJ...911..102O}
---. 2021, ApJ, 911, 102, \dodoi{10.3847/1538-4357/abeb6e}

\bibitem[{{Planck Collaboration} {et~al.}(2020){Planck Collaboration},
  {Aghanim}, {Akrami}, {Ashdown}, {Aumont}, {Baccigalupi}, {Ballardini},
  {Banday}, {Barreiro}, {Bartolo}, {Basak}, {Battye}, {Benabed}, {Bernard},
  {Bersanelli}, {Bielewicz}, {Bock}, {Bond}, {Borrill}, {Bouchet}, {Boulanger},
  {Bucher}, {Burigana}, {Butler}, {Calabrese}, {Cardoso}, {Carron},
  {Challinor}, {Chiang}, {Chluba}, {Colombo}, {Combet}, {Contreras}, {Crill},
  {Cuttaia}, {de Bernardis}, {de Zotti}, {Delabrouille}, {Delouis}, {Di
  Valentino}, {Diego}, {Dor{\'e}}, {Douspis}, {Ducout}, {Dupac}, {Dusini},
  {Efstathiou}, {Elsner}, {En{\ss}lin}, {Eriksen}, {Fantaye}, {Farhang},
  {Fergusson}, {Fernandez-Cobos}, {Finelli}, {Forastieri}, {Frailis},
  {Fraisse}, {Franceschi}, {Frolov}, {Galeotta}, {Galli}, {Ganga},
  {G{\'e}nova-Santos}, {Gerbino}, {Ghosh}, {Gonz{\'a}lez-Nuevo}, {G{\'o}rski},
  {Gratton}, {Gruppuso}, {Gudmundsson}, {Hamann}, {Handley}, {Hansen},
  {Herranz}, {Hildebrandt}, {Hivon}, {Huang}, {Jaffe}, {Jones}, {Karakci},
  {Keih{\"a}nen}, {Keskitalo}, {Kiiveri}, {Kim}, {Kisner}, {Knox},
  {Krachmalnicoff}, {Kunz}, {Kurki-Suonio}, {Lagache}, {Lamarre}, {Lasenby},
  {Lattanzi}, {Lawrence}, {Le Jeune}, {Lemos}, {Lesgourgues}, {Levrier},
  {Lewis}, {Liguori}, {Lilje}, {Lilley}, {Lindholm}, {L{\'o}pez-Caniego},
  {Lubin}, {Ma}, {Mac{\'\i}as-P{\'e}rez}, {Maggio}, {Maino}, {Mandolesi},
  {Mangilli}, {Marcos-Caballero}, {Maris}, {Martin}, {Martinelli},
  {Mart{\'\i}nez-Gonz{\'a}lez}, {Matarrese}, {Mauri}, {McEwen}, {Meinhold},
  {Melchiorri}, {Mennella}, {Migliaccio}, {Millea}, {Mitra},
  {Miville-Desch{\^e}nes}, {Molinari}, {Montier}, {Morgante}, {Moss}, {Natoli},
  {N{\o}rgaard-Nielsen}, {Pagano}, {Paoletti}, {Partridge}, {Patanchon},
  {Peiris}, {Perrotta}, {Pettorino}, {Piacentini}, {Polastri}, {Polenta},
  {Puget}, {Rachen}, {Reinecke}, {Remazeilles}, {Renzi}, {Rocha}, {Rosset},
  {Roudier}, {Rubi{\~n}o-Mart{\'\i}n}, {Ruiz-Granados}, {Salvati}, {Sandri},
  {Savelainen}, {Scott}, {Shellard}, {Sirignano}, {Sirri}, {Spencer},
  {Sunyaev}, {Suur-Uski}, {Tauber}, {Tavagnacco}, {Tenti}, {Toffolatti},
  {Tomasi}, {Trombetti}, {Valenziano}, {Valiviita}, {Van Tent}, {Vibert},
  {Vielva}, {Villa}, {Vittorio}, {Wandelt}, {Wehus}, {White}, {White},
  {Zacchei}, \& {Zonca}}]{2020AA...641A...6P}
{Planck Collaboration}, {Aghanim}, N., {Akrami}, Y., {et~al.} 2020, A\&A, 641,
  A6, \dodoi{10.1051/0004-6361/201833910}

\bibitem[{{Pol} {et~al.}(2021){Pol}, {Burke-Spolaor}, {Hurley-Walker},
  {Blumer}, {Johnston}, {Keith}, {Keane}, {Burgay}, {Possenti}, {Petroff}, \&
  {Bhat}}]{2021ApJ...911..121P}
{Pol}, N., {Burke-Spolaor}, S., {Hurley-Walker}, N., {et~al.} 2021, ApJ, 911,
  121, \dodoi{10.3847/1538-4357/abe70d}

\bibitem[{Rickett(1990)}]{ric90}
Rickett, B.~J. 1990, araa, 28, 561

\bibitem[{{Schlafly} \& {Finkbeiner}(2011)}]{2011ApJ...737..103S}
{Schlafly}, E.~F., \& {Finkbeiner}, D.~P. 2011, ApJ, 737, 103,
  \dodoi{10.1088/0004-637X/737/2/103}

\bibitem[{{Schoen} {et~al.}(2021){Schoen}, {Leung}, {Masui}, {Michilli},
  {Chawla}, {Pearlman}, {Shin}, {Stock}, \& {CHIME/FRB
  Collaboration}}]{2021RNAAS...5..271S}
{Schoen}, E., {Leung}, C., {Masui}, K., {et~al.} 2021, Research Notes of the
  American Astronomical Society, 5, 271, \dodoi{10.3847/2515-5172/ac3af9}

\bibitem[{{Simard} \& {Ravi}(2021)}]{2021arXiv210711334S}
{Simard}, D., \& {Ravi}, V. 2021, arXiv e-prints, arXiv:2107.11334.
\newblock \doarXiv{2107.11334}

\bibitem[{{Tendulkar} {et~al.}(2017){Tendulkar}, {Bassa}, {Cordes}, {Bower},
  {Law}, {Chatterjee}, {Adams}, {Bogdanov}, {Burke-Spolaor}, {Butler},
  {Demorest}, {Hessels}, {Kaspi}, {Lazio}, {Maddox}, {Marcote}, {McLaughlin},
  {Paragi}, {Ransom}, {Scholz}, {Seymour}, {Spitler}, {van Langevelde}, \&
  {Wharton}}]{2017ApJ...834L...7T}
{Tendulkar}, S.~P., {Bassa}, C.~G., {Cordes}, J.~M., {et~al.} 2017, ApJL, 834,
  L7, \dodoi{10.3847/2041-8213/834/2/L7}

\bibitem[{{Tendulkar} {et~al.}(2021){Tendulkar}, {Gil de Paz}, {Kirichenko},
  {Hessels}, {Bhardwaj}, {{\'A}vila}, {Bassa}, {Chawla}, {Fonseca}, {Kaspi},
  {Keimpema}, {Kirsten}, {Lazio}, {Marcote}, {Masui}, {Nimmo}, {Paragi},
  {Rahman}, {Pay{\'a}}, {Scholz}, \& {Stairs}}]{2021ApJ...908L..12T}
{Tendulkar}, S.~P., {Gil de Paz}, A., {Kirichenko}, A.~Y., {et~al.} 2021, ApJL,
  908, L12, \dodoi{10.3847/2041-8213/abdb38}

\bibitem[{{Yao} {et~al.}(2017){Yao}, {Manchester}, \&
  {Wang}}]{2017ApJ...835...29Y}
{Yao}, J.~M., {Manchester}, R.~N., \& {Wang}, N. 2017, ApJ, 835, 29,
  \dodoi{10.3847/1538-4357/835/1/29}

\bibitem[{{Zhu} {et~al.}(2020){Zhu}, {Li}, {Luo}, {Miao}, {Zhang}, {Spitler},
  {Lorimer}, {Kramer}, {Champion}, {Yue}, {Cameron}, {Cruces}, {Duan}, {Feng},
  {Han}, {Hobbs}, {Niu}, {Niu}, {Pan}, {Qian}, {Shi}, {Tang}, {Wang}, {Wang},
  {Yuan}, {Zhang}, {Zhang}, {Cao}, {Feng}, {Gan}, {Gao}, {Gu}, {Guo}, {Hao},
  {Huang}, {Huang}, {Jiang}, {Jin}, {Li}, {Li}, {Li}, {Liu}, {Pan}, {Peng},
  {Qian}, {Shi}, {Song}, {Song}, {Sun}, {Sun}, {Wang}, {Wang}, {Wang}, {Xie},
  {Yan}, {Yang}, {Yang}, {Yao}, {Yu}, {Yu}, {Zhang}, {Zhang}, {Zhang}, {Zheng},
  {Zhou}, {Zhu}, {Zhu}, {Zhu}, {Zhu}, \& {Zhu}}]{2020ApJ...895L...6Z}
{Zhu}, W., {Li}, D., {Luo}, R., {et~al.} 2020, ApJL, 895, L6,
  \dodoi{10.3847/2041-8213/ab8e46}

\end{thebibliography}

\listofchanges{}

\end{document}